\documentclass[10pt,final,journal]{IEEEtran}
\usepackage[ruled,vlined,linesnumbered,resetcount]{algorithm2e}
\usepackage{graphicx,epstopdf,bm,amsmath,marvosym,nopageno,mathtools,amssymb,color,url,environ,array,tabularx}
\usepackage{environ,enumerate,multirow}
\renewcommand{\theenumi}{\alph{enumi}}

\graphicspath{{./figs/}}

\renewcommand{\v}[1]{\boldsymbol{\mathbf{#1}}} 
\newcommand{\opn}[1]{\operatorname{#1}} 
\newcommand{\card}[1]{\left|\mathbb{#1}\right|} 
\newcommand{\set}[1]{\mathbb{#1}} 
\newcommand{\norm}[2]{\lVert #1 \rVert_{#2}} 
\newcommand{\braces}[1]{\left\{#1\right\}} 
\newcommand{\tm}[0]{\times} 
\newcommand{\prt}[1]{\left(#1\right)} 
\newcommand{\sqb}[1]{\left[#1\right]} 
\newcommand{\abs}[1]{\left\lvert #1 \right\rvert} 

\newcommand{\inc}[1]{\in \set{C}^{#1}} 

\NewEnviron{meq}{
    \begin{align}
        \begin{split}
            \BODY
        \end{split}
    \end{align}
    }
\newtheorem{theorem}{Theorem}
\newtheorem{definition}{Definition}

\begin{document}

\title{Direction of Arrival Estimation with Sparse Subarrays}

\author{Wesley S. Leite, Rodrigo C. de Lamare, Yuriy Zakharov, Wei Liu and Martin Haardt \vspace{-2.5em}

	\thanks{This work was supported by FAPERJ, CNPq and FAPESP. The work of Y. Zakharov was supported in part by U.K. EPSRC under Grant P/V009591/1. W. S. Leite, and R. C. de Lamare are with Department of Electrical Engineering, PUC-Rio, Rio de Janeiro 22451-900, Brazil. R. C. de Lamare and Y. Zakharov are with the School of Physics, Engineering and Technology, University of York, York, UK. W. Liu is with Department of Electrical and Electronic Engineering, The Hong Kong Polytechnic University, Kowloon, Hong Kong. Martin Haardt is with the Communications Research Laboratory, Ilmenau University of Technology, Germany (e-mail: wesleysouzaleite@gmail.com, delamare@puc-rio.br, yury.zakharov@york.ac.uk, wliu.eee@gmail.com, martin.haardt@tu-ilmenau.de.}
}

\maketitle

\begin{abstract}
	This paper proposes design techniques for partially-calibrated sparse linear subarrays and algorithms to perform direction-of-arrival (DOA) estimation. First, we introduce array architectures that incorporate two distinct array categories, namely type-I and type-II arrays. The former breaks down a known sparse linear geometry into as many pieces as we need, and the latter employs each subarray such as it fits a preplanned sparse linear geometry.  Moreover, we devise two Direction of Arrival (DOA) estimation algorithms that are suitable for partially-calibrated array scenarios within the coarray domain. The algorithms are capable of estimating a greater number of sources than the number of available physical sensors, while maintaining the hardware and computational complexity within practical limits for real-time implementation. To this end, we exploit the intersection of projections onto affine spaces by devising the Generalized Coarray Multiple Signal Classification (GCA-MUSIC) in conjunction with the estimation of a refined projection matrix related to the noise subspace, as proposed in the GCA root-MUSIC algorithm. An analysis is performed for the devised subarray configurations in terms of degrees of freedom, as well as the computation of the Cramèr-Rao Lower Bound for the utilized data model, in order to demonstrate the good performance of the proposed methods. Simulations assess the performance of the proposed design methods and algorithms against existing approaches.
\end{abstract}

\begin{IEEEkeywords}
	Direction-of-Arrival Estimation, Sparse Linear Array, MUSIC, Non-Uniform Linear Arrays
\end{IEEEkeywords}

\IEEEpeerreviewmaketitle

\section{Introduction}

\IEEEPARstart{O}{ver} the past decade, a primary focus in sensor array processing has been the advancement of increasingly accurate and efficient techniques for direction of arrival (DOA) estimation. This particular problem has its own fundamental importance and is strongly related to the fields of radar, sonar, wireless communications, seismology, and radio astronomy. Therefore, the problem is of great interest for several scientific research domains and practical applications exploiting sensor array technologies \cite{VanTrees2002,mmimo}.

\subsection{Prior and Related Works}

Recently, many research works have focused on the particular problem of underdetermined DOA estimation \cite{Pal2010-1,Pal2010,Liu2016,rdcoprime,misc22,emisc}. This problem consists of the determination of directions of more signal sources than the number of physical sensors in the array. Other works also focus on the reduction of hardware complexity, which encompasses hardware costs and energy-saving capabilities that are closely related to system operational costs and portability. The standard approach to DOA estimation employs a coherent sensor array, where the RF chains associated with sensors are connected to the same local oscillator and the intersensor distance is perfectly known (the array manifold is known), as well as the propagation channel mismatches. This approach is expensive and may be difficult to implement in many real-world situations, especially for large aperture arrays \cite{Rieken2004,Tirer2020}. 

A possible solution to this problem is to divide the whole array into subarrays and approach the task in one of two different ways: i) pre-process the subarray data separately and forward the processing results to a central processing unit (CPU - non-coherent processing); or ii) compensate for the unknown parameters that are needed for coherent processing by estimating them. For example, one can develop techniques to estimate the phase shifts between the steering vectors of the subarrays. Those phase shifts can be due to unknown intersubarray distances, lack of proper synchronization in time (non-coherent arrays), and unknown propagation channel mismatches (fades) \cite{See2004}.

In \cite{Tirer2021}, a new method based on a convex optimization formulation derived from a bilinear arrangement and a convex relaxation procedure was developed. The authors estimate the phase shifts between the subarrays and then use the corrected data model to estimate the DOAs using the well-known sparse signal processing formulation presented in \cite{Malioutov2005}. However, this method was developed for a single snapshot scenario and has a prohibitive computational complexity for most practical online processing scenarios. Moreover, in general it is not capable of estimating more sources than the number of  physical sensors. 

In \cite{Ma2009}, an approach for quasi-stationary signals was proposed, relying on the vectorization operator to build up a full-rank effective array manifold. This method was modified to deal with stationary signals through the introduction of a subspace-based algorithm known as Spatial Smoothing Multiple Signal Classification (SS-MUSIC) \cite{Pal2010, Pal2010-1} that performs spatial smoothing on the vectorized model, in the context of coherent (fully calibrated) arrays, and assumes uncorrelated sources. The boundaries of the algorithm performance were established later on in \cite{wang2017}.

In \cite{Rieken2004}, a decentralized, non-coherent DOA estimation algorithm was introduced based on the method of Projection onto Convex Sets (POCS) to compute a global estimate with the MUSIC algorithm. This strategy, subsequently referred to as G-MUSIC in Section~\ref{sec:results}, is a generalization of MUSIC. Notwithstanding its advantages, the algorithm possesses certain limitations. It was not explicitly engineered to deal with sparse arrays, although it remains compatible at some degree with such configurations. Moreover, its capability to estimate the number of sources is inherently constrained by the total number of sensors in each subarray through a linear relationship. In \cite{pesavento2018}, a sparse recovery approach for DOA estimation in partly calibrated arrays that uses mixed nuclear norm and $l_1$-norm minimization was introduced. Despite its superior performance, it relies on iterative optimization, which is costly for some applications.

More recently, two novel subarray algorithms have been presented in \cite{suleiman2018}. The first one relies on a non-convex optimization framework, which is very costly and necessitates the initiation from a feasible point. Furthermore, it does not guarantee convergence to a global minimum, thus making it a potentially unreliable strategy. In contrast, the second algorithm, involving a procedure labelled as the SPICE algorithm, uses a convex optimization with guaranteed convergence, although it is unsuitable for real-world scenarios. 

Thus, we note that there is a gap for efficient methods that can employ sparse arrays in partially-calibrated scenarios, while also addressing the underdetermined DOA estimation problem in a practical manner. The aim of this study is to present cost-effective approaches to this problem through sparse subarray design, as well as a signal processing strategy to estimate more sources than sensors in each subarray. Thus, we can devise cost-effective methods to estimate as many sources as possible using sparse subarrays, while reducing hardware and computational costs through a partially-calibrated scheme.

\subsection{Contributions}

In this work, we present design techniques for partially-calibrated sparse linear subarrays and algorithms to perform DOA estimation. First, we detail array architectures that incorporate two distinct array categories, namely type-I and type-II arrays, whose preliminary results appeared in \cite{leite2022}. The former breaks down a known sparse linear geometry into as many pieces as we need, and the latter employs each subarray such as it fits a preconceived sparse linear geometry.  Moreover, we devise two DOA estimation algorithms that are suitable for partially-calibrated array scenarios within the coarray domain. In particular, we exploit the intersection of projections onto affine spaces and derive the Generalized Coarray Multiple Signal Classification (GCA-MUSIC), whose preliminary results and description were reported in \cite{Leite2022-22}, and then we develop a root version of GCA-MUSIC that is denoted as the GCA root-MUSIC (GCA-rMUSIC) algorithm. The GCA-MUSIC and GCA-rMUSIC algorithms are suitable for scenarios within the coarray domain, allow the estimation of more sources than sensors in each subarray by handling the underdetermined case and do not rely on potentially costly iterative schemes or non-convex optimization. An analysis is performed for the proposed subarray configurations in terms of degrees of freedom. Moreover, we derive the Cramèr-Rao Lower Bound (CRLB) and its associated Fisher Information Matrices for the adopted data model. Simulations assess the performance of the proposed design methods and algorithms against existing approaches.

\emph{Paper structure}: In Section \ref{sec:systemModel}, the data model is introduced. In Section \ref{sec:spSubDesign}, the type-I and type-II sparse arrays are presented. In Section \ref{sec:propAlg}, the proposed GCA-MUSIC and GCA-rMUSIC algorithms are derived. A complexity analysis for both algorithms is performed in Section \ref{sec:analysis}, along with the properties of type-I and type-II arrays, as well as the CRLB. Numerical results in Section \ref{sec:results} show the algorithms' and arrays' ability in identifying the sources, whereas Section \ref{sec:conclusions} draws the conclusions.

\emph{Notation}: The symbols $\set{S}$, $a$, $\v{a}$ and $\v{A}$ refer to sets, scalars, column vectors and matrices, respectively. $[a]$ means the set $\{1,\ldots,a\}$. $|\set{A}|$ denotes the cardinality of the set $\set{A}$. The symbols $\mathcal{R}(\v{A})$ and $\mathcal{N}(\v{A})$ are the range and null space of matrix $\v{A}$, respectively. $\v{I}_M$ is the $M\times M$ identity matrix. $[a]$ means the set $\{1,\ldots,a\}$. $\opn{blkdiag}(\cdot)$ is the block diagonal matrix, whereas $\opn{colspan}(\v{A})$ represents the column space of $\v{A}$.

\section{System Model and Problem Statement}\label{sec:systemModel}
The received signal model for a multiple measurements scenario is given by
\begin{equation}\label{eq:data_model_param}
    \v{x}_{\set{S}}(t)=\Tilde{\v{V}}(\v{\theta})\Tilde{\v{H}}\v{s}(t)+\v{n}_{\set{S}}(t),~t\in\sqb{T}
\end{equation}
where
\begin{meq}
    \Tilde{\v{V}}(\v{\theta}) & = \sqb{\v{V}(\v{\theta}_1),\ldots,\v{V}(\v{\theta}_D)}\inc{N\times LD}\\
    \v{V}(\v{\theta}_d) & = \opn{blkdiag}\prt{\v{a}_1(\theta_d),\ldots,\v{a}_L(\theta_d)}\inc{N\times L}\\
    \Tilde{\v{H}} & = \opn{blkdiag}\prt{\v{h}_1,\ldots,\v{h}_D}\inc{LD\times D}\\
    \v{h}_d & = \sqb{1,h_{2d},h_{3d},\ldots,h_{Ld}}^T\inc{L\times 1}
\end{meq}
for $l\in\{1,\cdots,L\}$ and $t\in\{1,\cdots,T\}$, where $L$ is the number of subarrays and $T$ is the number of snapshots. The vector $\v{x}_{\set{S}}(t)\inc{N}$ represents the array received signal, $\v{s}(t)\inc{D}$ is the source signal vector, and $\v{n}_{\set{S}}(t)\inc{N}$ is the array noise vector. There are $D$ impinging sources with normalized directions given by $\v{\theta}\in [-1,1)^{D}$ (sine of DOAs). The noise and the source signals are drawn from a circularly complex multivariate Gaussian distribution. The noise is spatially and temporally white, and the noise and sources are uncorrelated. The vector $\v{h}_{d}$ contains the calibration parameters associated with all the subarrays (first subarray as reference) for the $d$-th direction, and $\sqb{\v{a}_{l}(\theta)}_k=\exp(j\pi n_k\theta)$ is the $k$-th element of the steering vector of the $l$-th subarray, $n_k\in{\set{S}_l}$ is the $k$-th sensor position as an integer multiple of the minimum intersensor distance, the set of integers $\set{S}_l$ and $N_l$ define the sensor locations and number of sensor elements of the $l$-th subarray, respectively.

The received signal in (\ref{eq:data_model_param}) can alternatively be rewritten as 
\begin{meq}\label{eq:equiv_data_model_fin}
    \v{x}_{\set{S}}(t) & = \begin{bmatrix}\v{V}(\v{\theta}_1)\v{h}_1,\ldots,\v{V}(\v{\theta}_D)\v{h}_D\\ \end{bmatrix}\v{s}(t)+\v{n}_{\set{S}}(t)\\
                        & = \begin{bmatrix} 
 								\v{A}_1(\v{\theta})\\
 								\v{A}_2(\v{\theta})\v{G}_2\\
 								\vdots\\
 								\v{A}_L(\v{\theta})\v{G}_L\\
 							\end{bmatrix}\v{s}(t)+\v{n}_{\set{S}}(t)
\end{meq}
where $\v{A}_l(\v{\theta})\inc{N_l\tm D}$ is the $l$-th calibrated subarray manifold\footnote{Notice that the subarrays are calibrated; however, the whole array is only partially calibrated because the differences in calibration parameters between different subarrays are unknown (e.g., their relative distances).} with geometry defined by the set of integers $\set{S}_l$, and $\v{G}_l=\opn{diag}\prt{\sqb{h_{l1},\ldots,h_{lD}}}$. $\v{G}_1$ is the identity matrix of order $D$, which is why it is not included in (\ref{eq:equiv_data_model_fin}). The array received signal $\v{x}_{\set{S}}(t)$ can be partitioned as 
\begin{equation}\label{eq:subarray_rec_sig}
    \v{x}_{\set{S}}(t) = \begin{bmatrix}
                         	\v{x}_{\set{S}_1}^T(t),\v{x}_{\set{S}_2}^T(t),\ldots,\v{x}_{\set{S}_L}^T(t)\\
                         \end{bmatrix}^T
\end{equation}
From that, it is clear that each subarray received signal in (\ref{eq:subarray_rec_sig}) can be rewritten as
\begin{meq}
    \v{x}_{\set{S}_l}(t) & = \v{A}_l(\v{\theta})\v{G}_l\v{s}(t)+\v{n}_{\set{S}_l}(t)
\end{meq}
$\v{x}_{\set{S}_l}(t)\inc{N_l}$ is the $l$-th subarray received signal snapshot, and $\v{n}_{\set{S}_l}(t)$ is the subarray noise vector measurement\footnote{To simplify the equations, we assume equal SNRs across all subarrays, thereby standardizing the noise levels. Considering subarrays with differing SNRs represents a potential avenue for future research.}. If we calculate the covariance matrix for each of the subarray data, we have
\begin{meq}\label{eq:cov_mat_sub}
    \v{R}_{\set{S}_l} & = \v{A}_l(\v{\theta})\v{G}_l \v{R}_s\v{G}_l^H\v{A}_l^H(\v{\theta})+\sigma_n^2\v{I}_{N_l}\\
    & = \v{A}_l(\v{\theta})\v{R}_s\v{A}_l^H(\v{\theta})+\sigma_n^2\v{I}_{N_l}\\
\end{meq}
where $\sigma_n^2$ is the noise variance, $\v{R}_{\set{S}_l}$ is the received signal covariance matrix of the $l$-th subarray. In (\ref{eq:cov_mat_sub}) we have used the fact that each calibration parameter $h_{ld}$ has a unitary module, according to what follows:  
\begin{meq}\label{eq:cov_subarrays}
    \v{G}_l \v{R}_s\v{G}_l^H & = \opn{diag}\prt{h_{l1} p_1 h_{l1}^{*},\ldots,h_{lD} p_1 h_{lD}^{*}}\\
    & = \opn{diag}\prt{|h_{l1}|^2 p_1,\ldots,|h_{lD}|^2 p_1}\\
    & = \opn{diag}(\v{p})\\
    & = \v{R}_s 
\end{meq}
where $\v{p}=\sqb{\sigma_1^2,\ldots,\sigma_D^2}$ contains the powers of the sources.

Remark$~1$: This demonstrates that, although the subarrays are uncalibrated relative to each other, the covariance matrix of each subarray does not depend on the calibration parameters. Conversely, while the subarrays are individually calibrated, the full array remains uncalibrated. We refer to this condition as a ``partially-calibrated array''.

By vectorizing (\ref{eq:cov_mat_sub}), we arrive at
\begin{equation}\label{eq:vec_op}
  \v{z}_{\set{S}_l} = \prt{\v{A}_{\set{S}_l}^{\ast}(\v{\theta})\circ\v{A}_{\set{S}_l}(\v{\theta})}\v{p}+\sigma_n^2\bar{\v{i}},
\end{equation}
where $\circ$ denotes the Khatri-Rao product, $\bar{\v{i}}=\sqb{\v{e}_1^T,\ldots,\v{e}_{N_l}^T}^T$ is the vectorization of the identity matrix, and $\v{e}_i$ is the $i$-th canonical basis vector (all-zeros vector except for a 1 at the $i$-th position). To simplify the equations, we will assume $N/L$ sensors for each subarray, where $N=\sum_{l=1}^L N_l$ is the total number of sensors. By removing the repeated rows in $\v{A}_{\set{S}_l}^{\ast}(\v{\theta})\circ\v{A}_{\set{S}_l}(\v{\theta})$ after their first occurrence (mirroring the operation in $\v{z}_{\set{S}_l}$ and $\bar{\v{i}}$)\footnote{In practice, with finite data sample, we average the repeated elements in $\v{z}_{\set{S}_l}$ instead of keeping only the first element. This gives a more accurate estimate of the statistic for this specific covariance lag.} and sorting the virtual sensor (coarray) elements in ascending order, we have
\begin{equation}\label{eq:scaa}
  \v{x}_{\set{D}_l}=\v{A}_{\set{D}_l}(\v{\theta})\v{p}+\sigma_n^2\v{i},
\end{equation}
where $\set{D}_l$ denotes the difference coarray set associated with the $l$-th subarray, $\sqb{\v{A}_{\set{D}_l}(\v{\theta})}_{m,d}=\exp(j\pi n_m \theta_d)$ with $n_m\in \set{D}_l$, $\v{x}_{\set{D}_l}\inc{|\set{D}_l|}$ is the $l$-th coarray received signal and $\v{i}\in\{0,1\}^{|\set{D}_l|}$ is an all-zero vector with the exception of a $1$ in its half position (element $(|\set{D}_l|+1)/2$). 

The problem of interest consists in estimating the source directions $\v{\theta}$ from the pre-processed data in model (\ref{eq:scaa}).  

\section{Sparse Subarray Design}\label{sec:spSubDesign}

There are many ways to perform the design of a sensor array that consists of subarrays. For example, one could rely on the split of a hypothetical geometry into smaller parts such that the aperture of the original array is greater than or equal to the sum of the apertures of the subarrays. The inequality instead of equality only comes from the fact that the subarrays are separated by a distance of at least one unit of intersensor spacing. This is called a type-I subarray that is formally defined next. 

\begin{definition}[Type-I Sparse Linear Array]
  A type-I Sparse Linear Array (type-I SLA) corresponds to an array of predefined sparse linear geometry $\set{S}$. The subsets defining the subarray geometries are generated from partitions $\set{S}_l\subset\set{S}$ such that if $s\in \set{S}_i$ and $f\in \set{S}_j$, $i<j$, then $s<f$.
\end{definition}

The second most intuitive way of generating an array from subarrays would be to take subarrays with a predefined geometry and combine them to generate an array of a given aperture. This configuration is what we refer to as a type-II array. Type-II arrays are assumed to exhibit the same geometric pattern, i.e., the parent array consists of a union of subarrays with the same geometry. In what follows, we establish the formal definition.

\begin{definition}[Type-II Sparse Linear Array]\label{def:IISLA}
  A type-II Sparse Linear Array (type-II SLA) corresponds to a union of subarrays with predefined sparse linear geometries $\set{S}_l$ (partitions of the array geometry $\set{S}$). The set $\set{S}_1$ defines the reference subarray. The remaining subarrays are derived from $\set{S}_1$ through $\set{S}_i=\set{S}_{i-1}^{\Delta_{i-1}}$, where $\Delta_{i-1}$ is a translation factor for all the elements of $\set{S}_{i-1}$ and is given by $\Delta_{i-1}=\mu+\kappa_{i-1}$, where $\mu$ is the normalized distance between subarrays (in terms of integer multiples of $d$, that represents the minimum intersensor spacing) and $\kappa_{i-1}$ is the aperture of the $(i-1)$-th subarray.  
\end{definition}

Since all subarrays are translated versions of the reference subarray, characterized only by a linear translation factor, it follows that for any $i=2,\ldots,L$, the relationship $\kappa_{i-1}=\kappa_{i}$ holds in this case, indicating that the subarrays share identical sensors placement, except for a translation factor. However, we highlight that the definition can be easily extended for non-identical subarrays.

In what follows, we consider that the sets $\set{S}$ and $\set{S}_l$ have cardinality $N$ and $N_l$, respectively, i.e., $\card{S}=N$ and $|\set{S}_l|=N_l$ and also $N=\sum_{l=1}^{L} N_l$. The definitions in both cases restrict the subarrays such that they obey the following rules: i) they do not share any sensors (partitions are pairwise disjoint); ii) they are colinear; and iii) the array is given by the union of the subarrays $\prt{\set{S}=\bigcup_{l=1}^L \set{S}_l}$. Note that the constraint  $N_l=N/L$ (i.e., $\kappa_1=\ldots=\kappa_L=\kappa$) $\forall l\in\sqb{L}$, implies that the subarrays used to generate type-II SLAs are multiple invariant \cite{Swindlehurst2001}.

The key difference between these sparse array definitions is that for type-I SLAs the subarrays are generated by splitting an array with a predefined sparse geometry. On the other hand, for type-II SLA, the array is generated by a union of sparse linear subarrays with a predefined geometry.

Next, we define the difference coarray set, as well some other definitions that are important for the discussion that follows.

\begin{definition} (Difference Coarray Set - $\set{D}$) The difference coarray set represented by $\set{D}$ is a set associated with the sensor positions $\set{S}$ through
    \begin{equation}\label{eq:diffCoarraySet}
      \set{D} \triangleq \braces{n_1-n_2\mid (n_1,n_2)\in \set{S}^2}
    \end{equation}
arranged in ascending order and without repetition of elements.
\end{definition}

\begin{definition} (Degrees of Freedom - $\opn{DoF}$) The number of DoF of a geometry specified by $\set{S}$ is the cardinality of its difference coarray set. Then,
    \begin{equation}\label{eq:DoF}
      \opn{DoF} \triangleq \card{D}
    \end{equation}
\end{definition}

\begin{definition} (Central Difference Coarray Consecutive Set - $\set{U}$)
    The central difference coarray consecutive set, denoted by $\set{U}$, associated with a difference coarray set $\set{D}$ is the set
    \begin{equation}
        \set{U}=\braces{0}\cup\braces{\pm \sqb{m}}
    \end{equation}
    arranged in ascending order and without any repetition of elements, such that $m$ is chosen to maximize the cardinality of $\set{U}$ with $\set{U}\subseteq \set{D}$.
\end{definition}

\begin{definition} (Uniform Degrees of Freedom - $\opn{UDoF}$) The number of Uniform $\opn{DoF}$, denoted by $\opn{UDoF}$, of a geometry specified by $\set{S}$, is the cardinality of its central difference coarray consecutive set $(\set{U})$, i.e.,
    \begin{equation}\label{eq:uDoF}
      \opn{UDoF} \triangleq \card{U}
    \end{equation}
\end{definition}

Let us now consider a well-known sparse linear geometry: the minimum redundancy array (MRA) \cite{Moffet1968}. The MRA is designed to minimize the number of pairs of physical sensors with the same position difference under the restriction that the associated difference coarray set is \emph{hole-free}, which means that there are no missing lags in its covariance matrix. 

\begin{figure*}
  \centering
  \includegraphics[width=.7\textwidth]{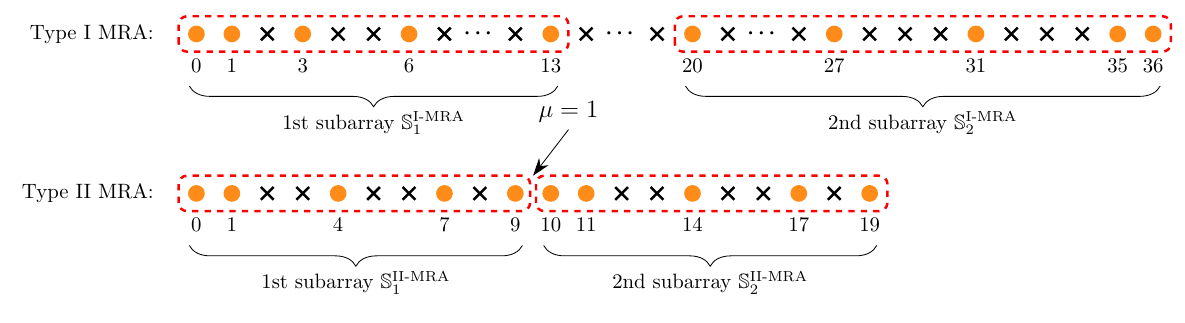}
  \vspace{-1.5em}\caption{Type-I and type-II MRA. Notice that the type-II array was generated with a parameter $\mu=1$ (spacing between subarrays).} \vspace{-1.5em} \label{fig:mra_example}
\end{figure*}

To illustrate the concepts of type-I and type-II arrays, consider the sparse geometry defined by the set
$\set{S}^{\text{MRA}} = \{0,1,3,6,13,20,27,31,35,36\}$, which corresponds to a MRA with $N=10$ sensors. We adopt $L=2$ subarrays. The corresponding type-I array is the original array itself, i.e., $\set{S}^{\text{I-MRA}}=\set{S}^{\text{MRA}}$, for an intersubarray displacement equal to $\mu=(20-13)=7$ and its subarrays are given by $\set{S}_{1}^{\text{I-MRA}}=\{0,1,3,6,13\}$ and $\set{S}_{2}^{\text{I-MRA}}=\{20,27,31,35,36\}$. On the other hand, consider two MRAs with $N=5$ sensors each given by $\set{S}_{1}^{\text{II-MRA}}=\{0,1,4,7,9\}$ and $\set{S}_{2}^{\text{II-MRA}}=\set{S}_{1}^{\Delta_1;\text{II-MRA}}=\{10,11,14,17,19\}$, with translation factor $\Delta_1=1+(9-0)=10$. Thus, $\set{S}^{\text{MRA}}=\set{S}^{\text{I-MRA}}\neq \set{S}^{\text{II-MRA}}$. For didactic purposes, Figure~\ref{fig:mra_example} shows a graphical representation of arrays in this example.

Before we proceed, it is worth discussing two additional sparse linear geometries, because they will be referred to later in the text: the two-level nested array (NAQ2) \cite{Pal2010} and the Second-Order Super Nested Array (SNAQ2) \cite{Liu2016}.

The so-called $n$-th level nested array (NAQn) was first introduced in \cite{Pal2010}. The discussion is restricted here to the two-level nested array (NAQ2). The design rule (positioning procedure) described by $\set{S}$ is given by 
 \begin{meq}\label{eq:na2}
    \mathbb{S} & =\{n,~n=0,1,\ldots,N_1-1\}\cup\\
               & \{n(N_1+1)-1,~n=1,2,\ldots,N_2\}
\end{meq}

The total number of sensors is $N=N_1+N_2$. Two of its fundamental properties are: a) hole free difference coarray ($\set{D}=\set{U}$); and b) the DoF/UDoF are given by $\opn{DoF}=\opn{UDoF}=2N_2(N_1+1)-1$. From that, notice that there are $\mathcal{O}(N_1N_2)$ available DoF for arrays with only $N_1+N_2$ sensors. The fact that NAQ2 structures have no holes in its coarray is a desired property for covariance matrix based algorithms, especially when spatial smoothing techniques are employed. The geometry  SNAQ2 has the same filled coarray as NAQ2 with the advantage of reduced mutual electromagnetic coupling. Lastly, we also introduce the so-called weight function.

\begin{definition} [Weight Function - $w(m)$]\label{def:weight_func}
  The weight function is an application
  \begin{equation}
    w: \set{Z} \to \braces{0}\cup\braces{\pm\sqb{N}}
  \end{equation}
  such that
  \begin{equation}
    w(m)=
    \begin{dcases}
      |\set{M}(m)| & \text{if $m\in \set{D}$} \\
      0            & \text{otherwise}         \\
    \end{dcases}
  \end{equation}
  where $\set{M}(m)=\braces{(n_1,n_2)\in \set{S}^2\mid n_1-n_2=m}$, i.e., $\set{M}(m)$ is a function that counts the number of sensor pairs with separation $m$ (covariance lag).
\end{definition}

\section{Proposed DOA Estimation Algorithms}\label{sec:propAlg}

In this section, we present two algorithms for DOA estimation using partially-calibrated arrays. First, we derive the spectrum-based Generalized Coarray Multiple Signal Classification (GCA-MUSIC), followed by its root-version, termed Generalized Coarray Root Multiple Signal Classification (GCA-rMUSIC). Both algorithms rely on the subspace properties of spatial smoothing performed on subcoarrays and on mixing rules for combining the mixed subspaces. We remark that other subspace-type algorithms \cite{jidf,jio,wljio,jiodoa,mcg,saalt,l1stap,l1cgstap,rdrbm,sbstap,okspme,lrcc,mskaesprit} can be developed to solve this problem.

\subsection{Generalized Coarray MUSIC Algorithm}

The proposed GCA-MUSIC aims to tackle the case of DoA estimation with partly calibrated sparse subarray geometries. This estimation procedure has many advantages over other techniques presented in the literature: it is capable of exploiting half of the DOF of the subcoarrays, presents super-resolution performance capabilities and has a reasonable trade-off between performance and computational cost. To begin with, we introduce the following definitions:

\begin{definition}[Sparse Subarray (SpSub)]
  The sparse subarrays are each defined as a partially calibrated part of the entire array. Within each sparse subarray, the sensors are coherent; that is, the sampling process is conducted using local oscillators that are synchronized in both frequency and phase, and the relative positions of the sensors within the subarray are known. They are denoted by $\set{S}_l$, with $l=1,\ldots,L$.
\end{definition}
\begin{definition}[Subcoarray (SCA)]
  A subcoarray is defined as the Difference Coarray associated with a given SpSub. They are denoted by $\set{D}_l$, with \emph{$l=1,\ldots,L$}.
\end{definition}
\begin{definition}[Spatially Smoothed Subcoarray (SS-SCA)]
  A spatially smoothed subcoarray is a SCA with a reduced dimension, which is determined by the parameter choices of a procedure of spatial smoothing. They are denoted by \emph{$\set{D}_l^i$}, with \emph{$l=1,\ldots,L$} and $i=1,\ldots, M$, where $M$ is both the aperture of each SS-SCA and its quantity.
\end{definition}

Notice that each SCA (associated to a specific SpSub), generated according to the mathematical procedure described by (\ref{eq:cov_mat_sub}), (\ref{eq:vec_op}) and (\ref{eq:scaa}), will have a total of $M$ SS-SCA. Then, we have a total of $M\cdot L$ SS-SCA for the whole array.

We consider the SpSub and its respective SCA with central contiguous part (virtual ULA) large enough to allow the recovery of all DOAs of the sources. To simplify the equations, we will assume that the SpSub has a filled SCA (no holes in virtual domain), i.e., the second-order statistics associated to each SpSub contain all the correlation lags from 0 up to $\kappa = (|\set{D}_l|-1)/2 = (\text{sDoF}-1)/2$, where sDoF is the number of degrees of freedom for each SCA. 

By resorting to the rank properties, it is clear that the outer product $\v{x}_{\set{D}_l}\v{x}_{\set{D}_l}^H$ in (\ref{eq:scaa}) is rank deficient. Then, we build up this rank using $M=\kappa+1=(\text{sDoF}+1)/2$ SS-SCA (forward spatial smoothing), for each SCA, according to
\begin{equation}\label{eq:ss_eq}
  \v{R}_{\set{D}_l}^{\text{SS}} =\frac{1}{M}\sum_{i=1}^{M}\v{x}_{\set{D}_l^i}\prt{\v{x}_{\set{D}_l^i}}^H
\end{equation}
where $\v{x}_{\set{D}_l^i}=\v{J}_i\v{x}_{\mathbb{D}_l}\inc{M}$ is the $i$-th overlapping SS-SCA of the $l$-th SCA, $\v{J}_i = \sqb{\v{0}_{M\times (M-i)},\v{I}_M,\v{0}_{M\times(i-1)}}$ is a selection matrix, starting ($i=1$) from the maximum value of the contiguous part of the SCA. Each $\v{R}_{\set{D}_l}^{\text{SS}}$, originated from each of the partially calibrated subarrays provides rough estimates of the sources DoA. The second problem we address is how to integrate the processing to benefit from the estimates of each SpSub in a cohesive manner. This integration of the spectrum is crucial for enhancing estimation performance, as will be demonstrated further.

To perform the signal decomposition, we adopt a similar strategy as described in \cite{Rieken2004}. The signal and noise subspaces of $\v{R}_{\set{D}_l}^{\text{SS}}$ can be obtained from the following eigenvalue decomposition (EVD):
\begin{equation}\label{eq:ss-sca-svd}
  \v{R}_{\set{D}_l}^{\text{SS}} =
  \begin{bmatrix}
    \v{U}_{l}, \v{V}_{l}
  \end{bmatrix}
  \opn{diag}\prt{\beta_{1}^l,\ldots,\beta_{M}^l}
  \begin{bmatrix}
    \v{U}_{l}^H \\ \v{V}_{l}^H
  \end{bmatrix}
\end{equation}
that has the same eigenvectors as those associated with a covariance matrix whose data were obtained with the first array manifold of the spatial smoothing procedure, denoted by $\v{A}_{\set{D}_l^1}(\v{\theta})$ (the virtual array manifold corresponding to the last $M$ rows of $\v{A}_{\set{D}_l}(\v{\theta})$). Then, we can state that  $\mathcal{R}\prt{\v{A}_{\set{D}_l^1}(\v{\theta})}=\mathcal{R}\prt{\v{U}_{l}}$ and $\mathcal{N}\prt{\v{A}_{\set{D}_l^1}^H(\v{\theta})}=\mathcal{R}\prt{\v{V}_{l}}$, where $\mathcal{R}(\cdot)$ and $\mathcal{N}(\cdot)$ denote the range and null spaces of a matrix. Particularly in this case, the subarrays are partially-calibrated and so the statistics are divided and must be processed separately up to some degree.

The $i$-th SS-SCA of the $l$-th SCA has its signal part contained in $\mathcal{R}\prt{\v{U}_{l}}$. Define
\begin{equation}
  \Tilde{\v{U}}_l = \opn{blkdiag}\prt{\v{I}_{M(l-1)},\v{U}_l,\v{I}_{M(L-l)}}
\end{equation}
and
\begin{equation}
  \Tilde{\v{V}}_l = \opn{blkdiag}\prt{\v{I}_{M(l-1)},\v{V}_l,\v{I}_{M(L-l)}}
\end{equation}
Clearly, $\mathcal{R}\prt{\v{U}_l} \subset \mathcal{R}\prt{\Tilde{\v{U}}_l}$, i.e., the mixed block-diagonal structure imposes that the range space of $\Tilde{\v{U}}_l$ is a superset of the corresponding signal subspace $\v{U}_l$. However, the signal subspace lies in $\mathcal{R}\prt{\Tilde{\v{U}}_l}~\forall~l\in\{1,\ldots,L\}$.

Every affine set is both unbounded and convex. Furthermore, each of the subspaces denoted by $\mathcal{R}\prt{\v{U}_l}$ is affine. Similarly, the corresponding noise subspaces, denoted by $\mathcal{R}\prt{\v{V}_l}$, are also affine. Consider the set
\begin{equation}
  \mathbb{W} = \bigcap_{l=1}^{L} \mathcal{R}\prt{\Tilde{\v{V}}_l}
\end{equation}
In order to project a point onto $\set{W}$ (intersection of noise subspaces), we can iterate through successive projections onto each of the $\mathcal{R}\prt{\Tilde{\v{V}}_l}$ by resorting to the method of projection onto convex sets (POCS) \cite{Feng2007,Xia2008,Rieken2004}. Since the matrices $\Tilde{\v{V}}_l$ are orthogonal, their projection matrices are given by
\begin{meq}
  \v{P}_{\Tilde{\v{V}}_l}  & = \Tilde{\v{V}}_l\Tilde{\v{V}}_l^H\\
  & = \opn{blkdiag}\prt{\v{I}_{M(l-1)},\v{V}_l\v{V}_l^H,\v{I}_{M(L-l)}}
\end{meq}
The successive projections lead us to
\begin{meq}\label{eq:overall_matrix}
  \v{P} & = \overset{\curvearrowright}{\prod^{L}_{l=1}} \v{P}_{\Tilde{\v{V}}_l}\\
  & = \opn{blkdiag}\prt{\v{V}_1\v{V}_1^H,\v{V}_2\v{V}_2^H,\ldots,\v{V}_L\v{V}_L^H}
\end{meq}
which is idempotent and thus represents the final matrix after the convergence of POCS. The notation  $\overset{\curvearrowright}{\prod^{L}_{l=1}} \v{A}_l$ means the matrix product $\v{A}_1\cdot \v{A}_2\cdot\ldots\cdot\v{A}_L$. Let us define the composed steering vector in the SS-SCA coarray domain. We can write
\begin{equation}\label{eq:steer_vec_comp}
  \v{a}_{\set{D}}^{\text{SS}} (\theta) = \sqb{\prt{\v{a}_{\set{D}_1}^{\text{SS}}(\theta)}^T,\prt{\v{a}_{\set{D}_2}^{\text{SS}}(\theta)}^T,\ldots,\prt{\v{a}_{\set{D}_L}^{\text{SS}}(\theta)}^T}^T
\end{equation}
where
\begin{equation}\label{eq:stee_vec_coarray}
  \v{a}_{\set{D}_l}^{\text{SS}}(\theta) = \sqb{1,e^{j\pi\theta},\ldots,e^{j\pi(M-1)\theta}}^T
\end{equation}
that corresponds to the steering vector of  $\v{A}_{\set{D}_l^1}(\v{\theta})$ and does not depend on the subarray index $l$. This comes from the fact that each of the SCA is a filled coarray ULA and has the same number of degrees of freedom $|\set{D}_1|=\ldots=|\set{D}_L|$.

The steering vector in (\ref{eq:steer_vec_comp}) is orthogonal to the matrix $\opn{blkdiag}\prt{\v{V}_1,\v{V}_2,\ldots,\v{V}_L}$. Then, the quadratic form built from the matrix $\v{P}$ and the composite steering vector in (\ref{eq:steer_vec_comp}) should have a null for the true directions $\theta$, similarly to the MUSIC procedure. From that, the pseudo-spectrum can be obtained through
\begin{meq}\label{eq:pseudo_spec}
  P(\theta_i^g) & = \frac{1}{\prt{\v{a}_{\set{D}}^{\text{SS}} (\theta)}^H\v{P}\prt{\v{a}_{\set{D}}^{\text{SS}}(\theta)}}\\
  & =\frac{1}{\sum_{l=1}^{L}\prt{\v{a}_{\set{D}_l}^{\text{SS}} (\theta_i^g)}^H\v{V}_l\v{V}_l^H\v{a}_{\set{D}_l}^{\text{SS}} (\theta_i^g)},~i\in[g]
\end{meq}
where $\v{\theta}^g\in (-1,1]^g$ is a grid, i.e., the discretization of the continuous search area of candidate DOA angles. The true DOAs are estimated as the angles on the grid that maximize (\ref{eq:pseudo_spec}). GCA-MUSIC is summarized in Algorithm~\ref{alg:gca_music}.
\begin{algorithm}[h!]
  \DontPrintSemicolon
  \SetKwInOut{Inp}{Input}
  \SetKwInOut{Out}{Output}
  \Inp{subarray geometries $\mathbb{S}_l$, data matrix for each subarray $\hat{\v{X}}_{\set{S}_l}$, grid $\v{\theta}^g\in (-1,1]^g$, number of sources $D$}
  \BlankLine
  \For{$l\leftarrow 1$ \KwTo $L$}{
  $\hat{\v{R}}_{{\mathbb{S}}_l}\leftarrow (1/T)\hat{\v{X}}_{\set{S}_l}\hat{\v{X}}_{\set{S}_l}^H$\;
  $\hat{\v{z}}_{\set{S}_l} \leftarrow \opn{vec}\prt{\hat{\v{R}}_{{\mathbb{S}}_l}}$\;
  Average the repeated elements in $\v{z}_l$ and sort the corresponding coarray locations to obtain $\hat{\v{x}}_{\set{D}_l}$\;
  $\hat{\v{R}}_{\set{D}_l}^{\text{SS}} \leftarrow \frac{1}{M}\sum_{i=1}^{M}\hat{\v{x}}_{\set{D}_l^i}\prt{\hat{\v{x}}_{\set{D}_l^i}}^H$\;
  Diagonalize $\hat{\v{R}}_{\set{D}_l}^{\text{SS}}$ through an EVD to find $\hat{\v{V}}_l$\;
  }
  \For{$i\leftarrow 1$ \KwTo $g$}{
  $\hat{P}(\theta_i^g)=\prt{\sum_{l=1}^{L}\prt{\v{a}_{\set{D}_l}^{\text{SS}} (\theta_i^g)}^H\hat{\v{V}}_l\hat{\v{V}}_l^H\v{a}_{\set{D}_l}^{\text{SS}} (\theta_i^g)}^{-1}$ \;
  }
  Find the peaks in $\hat{P}(\theta_i^g)$ and take the corresponding angles as the estimated DOAs\;
  \BlankLine
  \caption{GCA-MUSIC}\label{alg:gca_music}
  \Out{Estimated DOAs $\hat{\v{\theta}}$}
\end{algorithm}

Remark 2:
\begin{enumerate}
  \item The matrices $\v{R}_{\set{D}_l}^{\text{SS}}$ are orthogonally diagonalizable because of their structure as a sum of outer products (Hermitian symmetric);
  \item We stress the fact that for SS-SCA with the same number of DoF (virtual aperture) and filled coarray ULAs, the vectors $\v{a}_{\set{D}_l}^{\text{SS}}(\theta)$ do not depend on $l$, as stated in (\ref{eq:stee_vec_coarray}).
  \item The algorithm can indeed be extended for \emph{any} linear sparse subarray geometry, once it has the same physical orientation (not necessarily type-II arrays). Indeed, for algorithms with some SpSub that do not have a filled ULA in the coarray domain, the algorithm can be extended to employ only the central contiguous part of the coarray, i.e., the central part corresponding to the UDoF. In the simulations, an example will illustrate this case.
\end{enumerate}

Lastly, we would like to point out that GCA-MUSIC is capable of identifying more sources than sensors even for the partially calibrated array scenario. GCA-MUSIC can identify up to $(\text{sDoF}-1)/2$ sources. To write this quantity as a function of the number of physical sensors, we must define the geometry we are dealing with. For example, for sparse subarrays following a NAQ2 geometry, GCA-MUSIC can identify up to $(N/L)^2/2+N/(2L)-1$ sources, which is the same number of sources that the coarray MUSIC can identify with a coherent array of $N/L$ sensors with this geometry. 

It would be of great interest if we could identify more sources than the number of sensors in the \emph{whole} array. Under certain conditions, GCA-MUSIC can indeed identify more sources than the total number of sensors. For example, for the case of the NAQ2 geometry that we discussed earlier, if we have $L=3$ subarrays and $N>16$ in a type-II structure, then the number of sources that can be identified is greater than the number of sensors in the whole arrangement. For example, for $L=3$ and $N=18$, GCA-MUSIC allows the identification of up to $D=20$ sources. This requires a case by case analysis for other geometries.

Additionally, the subarray displacement $\mu$ does not have to be known and the subarrays do not need to be collinear. The only requirement is that each SCA has a central contiguous part ($\set{U}$) that allows the spatial smoothing to be performed on it. In conclusion, it is feasible to use GCA-MUSIC in a heterogeneous subarray scenario incorporating MRA, NAQ2, and other geometrical configurations. Despite the resulting variations in the dimensions of the steering vectors $\v{a}_{\set{D}_l}^{\text{SS}}(\theta)$ and the disparate apertures of each SS-SCA for individual subarrays, the algorithm maintains its capabilities.

\subsection{Generalized coarray root-MUSIC}\label{sec:gca_rmusic}

In order to develop a version of root-MUSIC for partially calibrated sparse arrays, we exploit the root-MUSIC polynomial and the GCA-MUSIC of the previous section. We first review the standard root-MUSIC algorithm in \cite{Barabell1983}.


The steering vector of a coherent ULA can be written as
\begin{equation}\label{eq:polynomial_steering_vector}
  \v{f}(z) = \sqb{1,z,z^2,\ldots,z^{N-1}}^T
\end{equation}
which is evaluated at $z=\exp\prt{j\pi\theta}$, assuming an array with $N$ physical sensors. It is widely known that the noise subspace has all their eigenvectors orthogonal to the steering vectors of the array manifold, i.e., $\mathcal{R}(\v{V})\perp \v{f}(z)$ where $\v{V}$ is a basis for the noise subspace of the coherent ULA and $\v{f}(z)=\v{a}_{\set{S}}(\theta)$ denotes the corresponding array steering vector for the direction $\theta$. The inner product between those two quantities should be zero. From this, we can write the polynomial:
\begin{align}
Q(z) & = \norm{\v{f}^H(z)\v{V}}{2}^2\nonumber\\
       				& =  \v{f}^H(z)\v{V}\prt{\v{f}^H(z)\v{V}}^H\nonumber\\
       				& = \prt{\v{f}^{\ast}(z)}^T\v{V}\v{V}^H\v{f}(z)\nonumber\\
       				& = \v{f}^T(z^{\ast})\v{V}\v{V}^H\v{f}(z)\nonumber\\
       				& = \v{f}^T(1/z)\v{V}\v{V}^H\v{f}(z)\nonumber\label{eq:last_eq}
\end{align}
which is the polynomial form for the coherent array associated with the projection matrix $\v{V}\v{V}^H$ and the last equality comes from the fact that $\abs{z}=1$. Notice that in this case, the array geometry must be a ULA. Following the orthogonality conditions, this polynomial must have a zero if evaluated at the true directions. This is the standard root-MUSIC algorithm.

Clearly, in the case of partially calibrated sparse subarrays, where only the second-order statistics of each subarray are available and we do not have access to the cross second-order statistics, a general polynomial must be assembled by resorting to extra tools with the constraint that the subarrays are sparse. 
As with GCA-MUSIC, here we deal with each of the SpSub after the spatial smoothing procedure as described in (\ref{eq:ss-sca-svd}). In this case, we will use each of the noise subspaces $\v{V}_l$ to obtain a more accurate estimate of the true noise subspace. The proposed procedure relies on the intuition that a more accurate polynomial would be obtained by using the denominator of (\ref{eq:pseudo_spec}). We assume that each SCA can identify all the sources in the coarray domain. 

Inspired by root-MUSIC, we first observe that a vector is in the nullspace of (\ref{eq:overall_matrix}) if and only if each of its partitions of appropriate size is in the nullspace of the corresponding block. Consider the partitioned steering vector defined as
\begin{equation}\label{eq:polynomial_steering_vector_coarray}
  \v{f}_{\set{D}}(z) = \sqb{\v{f}^T_1(z),\ldots,\v{f}^T_L(z)}^T \inc{L M_l}
\end{equation}
where
\begin{equation}\label{eq:steer_vecvec}
  \v{f}_l(z) = \sqb{1,z,z^2,\ldots,z^{M_l-1}}^T \inc{M_l}
\end{equation}
and $M_l=(\text{sDoF}+1)/2$. Since each component of (\ref{eq:polynomial_steering_vector_coarray}) is necessarily in the range space of (\ref{eq:overall_matrix}), then we can write 
\begin{meq}\label{eq:polynomial_steering_vector_final}
  Q_{\set{D}}(z) & = \v{f}_{\set{D}}^T(1/z)\v{P}\v{f}_{\set{D}}(z)\\ 
  & = \sum_{l=1}^{L}\v{f}_l^T(1/z)\v{V}_{l}\v{V}_{l}^{H}\v{f}_l(z)
\end{meq}
This global coarray polynomial can then be employed to find the true DOAs. Notice that when this substitution is performed, some mathematical implications must be considered. We briefly discuss them in what follows. 

We can build this polynomial because each of the SpSub becomes a virtual ULA in the coarray domain. Its degree is a function of the degrees of freedom of the subarrays. The global polynomial in (\ref{eq:polynomial_steering_vector_final}) has degree $n_Q=2M_i-2$, where $i=\opn{argmax}_{l\in \{1,\ldots,L\}} M_l$, the index corresponding to the subarray with largest UDoF. Clearly, the higher order coefficients are only affected by the subarrays with large virtual aperture. This observation allows us to further simplify (\ref{eq:polynomial_steering_vector_final}) and gain some insight regarding its structure. Based on this polynomial, we can build the global projection matrix as 
\begin{equation}\label{eq:tilde_p}
  \tilde{\v{P}} = \v{P}_{\v{V}_i}+\sum_{\substack{l=1, l\neq i}}^{L} \tilde{\v{P}}_{\v{V}_l}
\end{equation}
where 
\begin{equation}
    \tilde{\v{P}}_{\v{V}_l}=\begin{bmatrix} \v{0}_{(M_i-M_l)\times M_l} \\ \v{I}_{M_l} \end{bmatrix} \v{P}_{\v{V}_l} \begin{bmatrix} \v{I}_{M_l},\v{0}_{M_l\times (M_i-M_l)} \end{bmatrix}
\end{equation}
Then, we can state that $\tilde{\v{P}}$ is a global noise projection matrix, that accounts for the sum of polynomials and thus represents the noise subspace more accurately compared to the noise subspace of a single subarray. We point out that (\ref{eq:tilde_p}) can even be used in scenarios where the subarrays have different SNRs, by exploiting similar strategies used in antenna diversity techniques \cite{juboori2018}. It suffices to apply weights on the noise projections and use an optimization criterion to find the best possible subspace estimate. For the sake of clarity and ease of understanding, we shall confine our analysis to subarrays with the same SNR. By combining (\ref{eq:steer_vecvec}) and (\ref{eq:tilde_p}), we have
\begin{equation}
    Q_{\set{D}}(z) = \v{f}_i^T(1/z)\tilde{\v{P}}\v{f}_i(z),
\end{equation}
where $i=\opn{argmax}_{l\in \{1,\ldots,L\}} M_l$, as previously mentioned. After the root pruning process, only the $D$ roots inside and closest to the unit circle are considered.

We highlight that GCA-rMUSIC is capable of estimating more sources than the number of sensors in each subarray, as it significantly increases the polynomial degree. Moreover, while root-MUSIC requires the array to be fully synchronized, GCA-rMUSIC, summarized in Algorithm~\ref{alg:gca_music_avprojm}, can estimate directions based on a global estimate of the nullspace. 

\begin{algorithm}
  \DontPrintSemicolon
  \SetKwInOut{Inp}{Input}
  \SetKwInOut{Out}{Output}
  \Inp{subarray geometries $\mathbb{S}_l$, data matrix for each subarray $\hat{\v{X}}_{\set{S}_l}$, grid $\v{\theta}^g\in (-1,1]^g$, number of sources $D$}
  \BlankLine
  \For{$l\leftarrow 1$ \KwTo $L$}{
  $\hat{\v{R}}_{{\mathbb{S}}_l}\leftarrow (1/T)\hat{\v{X}}_{\set{S}_l}\hat{\v{X}}_{\set{S}_l}^H$\;
  $\v{z}_{l} \leftarrow \opn{vec}\prt{\hat{\v{R}}_{{\mathbb{S}}_l}}$\;
  Average the repeated elements in $\v{z}_l$ and sort the corresponding coarray locations to obtain $\v{x}_{\set{D}_l}$\;
  $\v{R}_{\set{D}_l}^{\text{SS}} \leftarrow \frac{1}{M}\sum_{i=1}^{M}\v{x}_{\set{D}_l^i}\prt{\v{x}_{\set{D}_l^i}}^H$\;
  Diagonalize $\v{R}_{\set{D}_l}^{\text{SS}}$ through an EVD to find $\v{V}_l$\;
  $\v{P}_{\v{V}_l} \leftarrow \v{V}_{l}\v{V}_{l}^{H}$\;
  }
  Search for $i=\opn{argmax}_{l\in \{1,\ldots,L\}} M_l$\;
  $\tilde{\v{P}} = \v{P}_{\v{V}_i}+\sum_{\substack{l=1 \\ l\neq i}}^{L} \tilde{\v{P}}_{\v{V}_l}$\;
  Find the roots of $Q_{\set{D}}(z) = \v{f}_i^T(1/z)\tilde{\v{P}}\v{f}_i(z)$\;
  Select the $D$ roots inside and closest to the unit circle as $\hat{z}_d$\;
  Estimate the DOAs according to $\hat{\theta} = \prt{\angle \hat{z}_d}/\pi$\;
  \BlankLine
  \caption{GCA-rMUSIC} \label{alg:gca_music_avprojm}
  \Out{Estimated DOAs $\hat{\v{\theta}}$}
\end{algorithm}

\section{Analysis}\label{sec:analysis}

In this section, we present an analysis of degrees of freedom of the proposed subarray geometries, an assessment of the computational cost of the proposed DOA estimation algorithms, an evaluation of their identifiability properties and derive a CRLB.

\subsection{Analysis of degrees of freedom}\label{sec:anadof}

A natural question that arises from the above introduced (sub)array geometries is related to the manifold structure, as well as the DoF for a difference coarray scenario \cite{Leite2021,Liu2016}. In this section, we shed light upon these two important aspects.

\subsubsection{Type-I}
clearly, for type-I arrays there is no \emph{a priori} analytical relation between the subarray manifolds, due to the fact that the geometries change dramatically between subarrays. Also, the number of DoF for the predefined geometry and its type-I counterpart is obviously the same, for the case in which we keep the original intersubarray displacement and both geometries coincide. Moreover, there is no general rule to predict the DoF associated to the subarrays. This requires an analysis case by case for the sparse geometry employed.

\subsubsection{Type-II}
some analytical relations can be straightforwardly derived. In this case, to simplify the equations, we will assume that the subarrays are multiple invariant, i.e., they attend the sufficient condition of having the same number of physical sensors $N/L$ and thus the same aperture. Using the array manifold of the whole array expressed in matrix form, we can write
\begin{equation}\label{eq:type_II_manif}
  \v{A}_{\set{S}}=[\v{A}_{\set{S}_1}^T,(\v{A}_{\set{S}_1}\v{\Lambda}^{\Delta})^T,\ldots,(\v{A}_{\set{S}_1}\v{\Lambda}^{(L-1)\Delta})^T]^T
\end{equation}
where we dropped the dependence of the manifold on the DOAs to simplify the notation. The matrix $\v{\Lambda}$ is defined as
\begin{equation}
  \v{\Lambda} = \opn{diag}\prt{e^{j\pi\theta_1},\ldots,e^{j\pi\theta_D}}
\end{equation}
The multiple exponents of $\v{\Lambda}$ in (\ref{eq:type_II_manif}) represent the subarray translations along the straight line and are assumed to be $\Delta=\mu+\kappa$ (see Definition~\ref{def:IISLA}).

In what follows, we establish our most important result \cite{leite2022}: the number of DoF for type-II arrays is upper-bounded by a function of the DoF of the subarrays (sDoF), which can be theoretically calculated for a variety of geometries. This result is quantitatively described in Theorem~\ref{thm:dofsdof}.

\begin{theorem}\label{thm:dofsdof}
  Consider a type-II array with geometry as defined in Definition~\ref{def:IISLA} with equal-aperture subarrays. If $1\leq \mu \leq \kappa$, then the number of \emph{DoF} of the array $\set{S}$ is upper-bounded by \emph{$L(\text{sDoF}-1)+2(L-1)\mu+1$}, where \emph{sDoF} is the number of \emph{DoF} for each subarray. If $\mu>\kappa$, then the number of \emph{DoF} is equal to \emph{$(2L-1)\text{sDoF}$}.

  \begin{IEEEproof} Let $c(n)$ be a discrete-valued function that assumes the value of 1 if there is a sensor at $n$ or 0 otherwise. This function for a type-II array is given by $c(n)=c_1(n)+\ldots+c_{L}(n)$. Since the subarrays are translated versions of the reference array, $c_l(n)=c_1(n-\Delta_l)$, where $\Delta_l=(l-1)\Delta$. The weight function associated to that array (counts the number of spatial correlations with lag $n$) is defined through \cite{Pal2010-1} $w(n) = c(n)\circledast c^{-}(n)$, where $c^{-}(n)$ is the time reversal version of $c(n)$ and $\circledast$ denotes the digital convolution between two sequences. Then, one can write
    \begin{meq}\label{eq:proof_theo}
      w(n) & = \prt{\sum_{i=1}^L c_i(n)} \circledast \prt{\sum_{j=1}^L c_j^{-}(n)}\\
      & = \sum_{i,j=1}^L c_1(n-(i-1)\Delta)\circledast c_{1}^{-}(n-(j-1)\Delta)\\
      & = \sum_{i,j=1}^L c_1(n)\circledast c_{1}^{-}(n)\circledast\delta(n-\Delta_i+\Delta_j)\\
      & = \sum_{i,j=1}^L w_1(n-\Delta_i+\Delta_j)\\
    \end{meq}
    The number of DoF is the cardinality of the support of $w(n)$. This support set has always an odd number of elements, given that $w(n)$ is an even function. Clearly, from (\ref{eq:proof_theo}), the weight function of the reference array $w_1(n)$ is repeated along the domain with displacement factors given by $\Delta_j-\Delta_i$. Since $|\opn{supp}(w_1(n))|=\text{sDoF}$ and for $\mu>\kappa$ there is no superposition of the weight functions of the subarrays with a different $\Delta_j-\Delta_i$, then $|\opn{supp}(w(n))|=(2L-1)|\opn{supp}(w_1(n))|\Rightarrow \text{DoF}=(2L-1)\text{sDoF}$. Note that there are $2L-1$ diagonals in a $L\tm L$ matrix representing all possible $\Delta_j-\Delta_i$. For $1\leq \mu\leq \kappa$, the superposition implies that the support set of $w(n)$ ranges from $-(L-1)\Delta-\kappa$ to $(L-1)\Delta+\kappa$. Since $\kappa=(\text{sDoF-1})/2$, the support set of $w(n)$ has a maximum number of elements equal to $2\sqb{(L-1)\Delta+\kappa}+1$ or $L(\text{sDoF}-1)+2(L-1)\mu+1$. The equality holds if the difference coarray of the subarrays has no holes (no missing lags), as it indeed happens for many geometries like NAQ2 and (restricted) MRAs. 
  \end{IEEEproof}
\end{theorem}

\subsection{Arithmetic complexity}\label{sec:arith_complexity}
In terms of computational complexity, we can count the number of arithmetic operations such as additions and multiplications for each algorithm as it is described in what follows. For GCA-MUSIC, we have:
\begin{enumerate}[a)]
  \item \textbf{Step S1: estimation of covariance matrix}. This step computes each of the subarrays' received signal covariance matrix by means of its sample covariance estimate. Since the outer products involve vectors of dimension $N_l=N/L$, then the number of required additions and multiplications is $L[(T-1)(N/L)^2]$ and $L[T(N/L)^2]$, respectively.
  \footnote{Notice that we consider the case where all the subarrays have the same number of physical sensors to simplify the final expressions.};
  \item \textbf{Step S2: estimation of spatially-smoothed coarray covariance matrix}. In this case, the coarrays outer products are obtained from vectors of dimension $M$ (see (\ref{eq:ss_eq})). Then, the number of additions and multiplications is $L\cdot[(T-1)M^2]$ and $L\cdot[TM^2]$, respectively;
  \item \textbf{Step S3: EVD of $\v{R}_{\set{D}_l}^{\text{SS}}$}. Since the decompositions involve a square matrix of dimension $M$, the EVD takes about $4M^3/3+\mathcal{O}(M^2)$ operations;
  \item \textbf{Step S4: noise subspace projection matrices}. The projection matrix $\v{P}_{\v{V}_l}=\v{V}_l\v{V}_l^H=\v{I}-\v{U}_l\v{U}_l^H$  requires $L\cdot (B-1)M^2$ additions and $LBM^2$ multiplications, as it involves $L$ matrices of size $M\times B$, where $B=M-D$;
  \item \textbf{Step S5: pseudo-spectrum calculation}. The pseudo-spectrum, for each point in the grid, is calculated through $\prt{\v{a}_{\set{D}_l}^{\text{SS}} (\theta_i^g)}^H\hat{\v{V}}_l\hat{\v{V}}_l^H\v{a}_{\set{D}_l}^{\text{SS}} (\theta_i^g)$. It involves $LBM^2-L$ additions and $L(B+1)M^2+LM$ multiplications for each of the points in the grid. For a grid search with $g$ points the complexity for the whole search is $g(LBM^2-L)$ additions and $g(L(B+1)M^2+LM)$ multiplications.
\end{enumerate}

The arithmetic complexity is summarized in Table~\ref{tab:gca_root_music_complexity}.
\begin{table}
  \caption{Arithmetic complexity analysis of GCA-MUSIC} \vspace{-1em} \label{tab:gca_root_music_complexity} 
  \centering
  \begin{tabular}{cc|c}
    \hline\hline
    \multirow{2}{*}{\textbf{step}} & \multicolumn{2}{c}{\textbf{type of operation}}                                                 \\
    \cline{2-3}
                                   & \textbf{additions}                                                  & \textbf{multiplications} \\
    \hline\hline
    S1                             & $L(T-1)(N/L)^2$                                                     & $LT(N/L)^2$              \\
    \hline
    S2                             & $L(T-1)M^2$                                                         & $LTM^2$                  \\
    \hline
    S3                             & \multicolumn{2}{c}{$4M^3/3+\mathcal{O}(M^2)$}                                                  \\
    \hline
    S4                             & $L (B-1)M^2$                                                        & $L BM^2$                 \\
    \hline
    S5                             & $g(LBM^2-L)$                                                         & $g(L(B+1)M^2+LM)$                  \\
    \hline
    TOTAL         & \multicolumn{2}{c}{$4M^3/3+\mathcal{O}(M^2)$}                            \\
    \hline\hline
  \end{tabular}
\end{table}

For GCA-rMUSIC, the steps S1-S4 are the same as those of GCA-MUSIC. For conciseness, we do not repeat them here. The additional steps are described in what follows.

\begin{enumerate}[a)]
   \item \textbf{Step S5: averaged projection matrix calculation}. The averaged projection matrix is calculated by (\ref{eq:polynomial_steering_vector}) and involves $(LB-1)M^2$ additions and $(LB+1)M^2$ multiplications.
   \item \textbf{Step S6: polynomial root-finding}. 
   This technique relies on the SVD of a square matrix (companion matrix) whose dimension coincides with the polynomial degree \cite{golub2012}. Then, since the coarray polynomial in (\ref{eq:polynomial_steering_vector_final}) has degree $2M-2$, this step takes about $32M^3/3+\mathcal{O}(M^2)$ operations.
\end{enumerate}

\subsection{Identifiability}

According to the analysis proposed in \cite{suleiman2018}, which establishes the identifiability conditions for non-coherent processing schemes, we can further investigate the properties of type-II arrays regarding identifiability. For that, we assume that the subarrays have an arbitrary number of sensors. The composed coarray manifold for type-II arrays is given by
\begin{equation}\label{eq:composed_coarray_manifold}
  \v{A}_{\set{D}}=\sqb{\v{A}_{\set{D}_1}^T,\cdots,\v{A}_{\set{D}_L}^T}^T
\end{equation}
The maximum number of identifiable sources is given by the sufficient condition $D\leq \lfloor \rho/2\rfloor$, where $\rho$ is the Kruskal rank of (\ref{eq:composed_coarray_manifold}). Let us analyze two main cases separately: subarrays with the same and a different number of sensors. For subarrays with the same number of sensors, $\rho_{\v{A}_{\set{D}}} = \rho_{\v{A}_{\set{D}_l}}$ and thus the SCA length determines the maximum number of identifiable sources. In this particular case, $M-1$.
On the other hand, if the subarrays have a different number of sensors, then the identifiability will be limited by the subarray with the largest aperture. In this case, $D\leq \lfloor \opn{max}_{l}(\rho_{\v{A}_{\set{D}_l}})/2\rfloor$. 

\subsection{Derivation of the CRLB for the partially-calibrated scheme}\label{sec:deriv_crlb}

In this section, we derive a CRLB for partially-calibrated arrays. Unlike the approach presented in \cite{suleiman2018}, our proposed CRLB assumes that the impinging source signals are spatially coherent across all subarrays. Additionally, the bound outlined in \cite{See2004} is not applicable under two specific conditions: 1) when the number of sources exceeds the total number of sensors across all subarrays, and 2) when the sources are assumed to be uncorrelated \emph{a priori}. Our proposed bound overcomes these limitations.




In order to derive the CRLB, we define the parameter vector as
\begin{equation}
    \v{\phi} = \sqb{\v{\theta}^T,\v{p}^T,\sigma_n^2,\v{\nu}_2^T,\ldots,\v{\nu}_L^T,\v{\eta}_2^T,\ldots,\v{\eta}_L^T}
\end{equation}
where
\begin{meq}
    \v{\nu}_l & = \mathfrak{Re}\left\{\sqb{\v{h}_{l,1},\ldots,\v{h}_{l,D}}^T\right\}\\
    \v{\eta}_l & = \mathfrak{Im}\left\{\sqb{\v{h}_{l,1},\ldots,\v{h}_{l,D}}^T\right\}\\
\end{meq}
Let us find the Fisher Information Matrix (FIM) associated with the estimation of the vector $\v{\phi}$, that
is defined as
\begin{equation}
    \text{FIM} = -E\left\{\frac{\partial^2\ln p(\v{x}_{\set{S}}(t);\v{\phi})}{\partial\phi_i\partial\phi_j}\right\},
\end{equation}
where $p(\v{x}_{\set{S}}(t);\v{\phi})$ is the probability density function of $\v{x}_{\set{S}}(t)$ which depends on the nonrandom parameter vector $\v{\phi}$. The covariance matrix of the data model is such that $x_{\set{S}}(t)\sim \mathcal{CN}(0,\v{R}_{\set{S}})$ and
\begin{equation}\label{eq:cov_mat_cal_param}
    \v{R}_{\set{S}} = \Tilde{\v{V}}(\v{\theta})\Tilde{\v{H}}\v{R}_s\Tilde{\v{H}}^H\Tilde{\v{V}}^H(\v{\theta})+\sigma_n^2\v{I} 
\end{equation}

The log-likelihood for a single snapshot is given by\cite{VanTrees2002}
\begin{equation}
\mathcal{L}_{\v{x}}(\v{\phi}) = -\ln |\pi \v{R}_{\set{S}}(\v{\phi})|-\v{x}^H\v{R}_{\set{S}}^{-1}(\v{\phi})\v{x}
\end{equation}
The elements of the FIM are given by
\begin{meq}\label{eq:gen_fim_formula}
    \sqb{\text{FIM}}_{ij} & = -E\sqb{\frac{\partial^2 \mathcal{L}_{\v{x}}(\v{\phi})}{\partial\phi_i\partial\phi_j}}\\
    & = T\opn{tr}\sqb{\v{R}_{\set{S}}^{-1}(\v{\phi})\frac{\partial \v{R}_{\set{S}}(\v{\phi})}{\partial \phi_i}\v{R}_{\set{S}}^{-1}(\v{\phi})\frac{\partial \v{R}_{\set{S}}(\v{\phi})}{\partial \phi_j}}\\
\end{meq}
Before we calculate the derivatives in (\ref{eq:gen_fim_formula}), we first divide the parameter vector into two blocks, as follows
\begin{equation}
    \v{\phi} = \begin{bmatrix}\v{\phi}_1^T & \v{\phi}_2^T\end{bmatrix}^T
\end{equation}
where $\v{\phi}_1=\v{\theta}$ and $\v{\phi}_2 = \sqb{\v{p}^T,\sigma_n^2,\v{\nu}_2^T,\ldots,\v{\nu}_L^T,\v{\eta}_2^T,\ldots,\v{\eta}_L^T}^T$. The FIM matrix can be partitioned as 
\begin{equation}
    \text{FIM} = \begin{bmatrix}
                  \text{FIM}_{\v{\phi}_1\v{\phi}_1}  & \text{FIM}_{\v{\phi}_1\v{\phi}_2}\\
                  \text{FIM}_{\v{\phi}_2\v{\phi}_1}  & \text{FIM}_{\v{\phi}_2\v{\phi}_2}
                 \end{bmatrix}   
\end{equation}
The CRLB matrix is defined as 
\begin{equation}
    \v{C}_{\text{CRLB}}(\v{\phi}) = \begin{bmatrix}
              \v{C}_{\text{CRLB}}(\v{\phi}_1\v{\phi}_1)  & \v{C}_{\text{CRLB}}(\v{\phi}_1\v{\phi}_2)\\
              \v{C}_{\text{CRLB}}(\v{\phi}_2\v{\phi}_1) & \v{C}_{\text{CRLB}}(\v{\phi}_2\v{\phi}_2)\\
              \end{bmatrix}
\end{equation}
Due to the fact that $\v{C}_{\text{CRLB}}(\v{\phi}) = \text{FIM}^{-1}$ and using the inverse formula for a block-partitioned matrix, we have
\begin{meq}\label{eq:crm_phi1_phi1}
\v{C}_{\text{CRLB}}(\v{\phi}_1\v{\phi}_1) = & \left[\text{FIM}_{\v{\phi}_1\v{\phi}_1}-\text{FIM}_{\v{\phi}_1\v{\phi}_2}\times\right.\\
&\left.\text{FIM}^{-1}_{\v{\phi}_2\v{\phi}_2}\text{FIM}_{\v{\phi}_2\v{\phi}_1}\right]^{-1}
\end{meq}
Since we are mainly interested in the estimation of the direction parameters, we are seeking the matrix $\v{C}_{\text{CRLB}}(\v{\phi}_1\v{\phi}_1)$. Let us proceed with the computation of each of the FIM blocks in (\ref{eq:crm_phi1_phi1}) using the expression in (\ref{eq:gen_fim_formula}) with the corresponding partial derivatives. 

In order to do so, we calculate the FIM matrices related to the variables in the parameter vector $\v{\phi}$, as well as their crossed FIMs. The key in the derivation process is to: 
\begin{itemize}
\item Compute the derivatives $\partial \v{R}_{\set{S}}(\v{\phi})/\partial \phi_i$;
\item Put the resulting expression in (\ref{eq:gen_fim_formula}) in a compact form. We proceed as follows:
\end{itemize}

\noindent
I) Computation of derivatives

\begin{itemize}

\item Derivative with respect to $\v{\theta}$:
\begin{meq}\label{eq:der_theta_initial}
    \frac{\partial \v{R}_{\set{S}}(\v{\phi})}{\partial \theta_i} = & \frac{\partial \Tilde{\v{V}}(\v{\theta})\Tilde{\v{H}}}{\partial \theta_i}\v{R}_s\Tilde{\v{H}}^H\Tilde{\v{V}}^H(\v{\theta})\\
    & + \Tilde{\v{V}}(\v{\theta})\Tilde{\v{H}}\v{R}_s\frac{\partial \Tilde{\v{H}}^H\Tilde{\v{V}}^H(\v{\theta})}{\partial \theta_i}
\end{meq}

The derivatives $\partial (\Tilde{\v{V}}(\v{\theta})\Tilde{\v{H}})/\partial \theta_i$ can be represented in compact form as 
\begin{equation}\label{eq:der_wrt_theta}
    \frac{\partial \Tilde{\v{V}}(\v{\theta})\Tilde{\v{H}}}{\partial \theta_i} = \v{D}\v{e}_i\v{e}_i^T
\end{equation}
where
\begin{equation}
\v{D} = \sqb{\frac{\partial \v{V}(\v{\theta}_1)\v{h}_1}{\partial \theta_1},\ldots,\frac{\partial \v{V}(\v{\theta}_D)\v{h}_D}{\partial \theta_D}}
\end{equation}
for which $\v{e}_i$ corresponds to the $i$-th canonical basis vector with all zeros except for a 1 in the $i$-th position. The goal of the outer product (projection) of the canonical basis vectors is to isolate the $i$-th column of $\v{D}$ and simultaneously nullify all its other columns. Notice that the parameters $\Tilde{\v{H}}$ are a function of the DOA sources in the general case. 

By substituting (\ref{eq:der_wrt_theta}) in (\ref{eq:der_theta_initial}), we have:
\begin{equation}\label{eq:deriv_theta}
    \frac{\partial \v{R}_{\set{S}}(\v{\phi})}{\partial \theta_i} = \v{D}\v{e}_i\v{e}_i^T\v{R}_s\v{W}^H+\v{W}\v{R}_s\v{e}_i\v{e}_i^T\v{D}^H
\end{equation}
where 
\begin{meq}\label{eq:w_definition}
    \v{W} & \triangleq \Tilde{\v{V}}(\v{\theta})\Tilde{\v{H}}\\
    & = \sqb{\v{V}(\v{\theta}_1)\v{h}_1,\ldots,\v{V}(\v{\theta}_D)\v{h}_D}\\    
\end{meq}
\begin{equation}
    \v{R}_s = \opn{diag}\prt{\v{p}}
\end{equation}

\item Derivative with respect to $\v{p}$ (source powers)
\begin{meq}\label{eq:deriv_sources_powers}
    \frac{\partial \v{R}_{\set{S}}(\v{\phi})}{\partial p_i} & = \v{W}\frac{\partial \v{R}_s}{\partial p_i}\v{W}^H\\
    & = \v{w}_i\v{w}_i^H\\
\end{meq}
where $\v{w}_i$ is the $i$-th column of $\v{W}$, defined in (\ref{eq:w_definition}).

\item Derivative with respect to $\sigma_n^2$

From (\ref{eq:cov_mat_cal_param}) the derivative of the covariance matrix with respect to the noise variance is given by
\begin{equation}\label{eq:deriv_noise_power}
    \frac{\partial \v{R}_{\set{S}}(\v{\phi})}{\partial \sigma_n^2} = \v{I}
\end{equation}

\item Derivatives with respect to the real part of the calibration parameters $\v{\nu}_l$
\begin{equation}\label{eq:der_nu_first}
    \frac{\partial \v{R}_{\set{S}}(\v{\phi})}{\partial \nu_{l,i}} = \frac{\partial \v{W}}{\partial \nu_{l,i}}\v{R}_s\v{W}^H+\v{W}\v{R}_s\frac{\partial \v{W}^H}{\partial \nu_{l,i}}
\end{equation}

The derivatives $\partial \v{W}/\partial \nu_{l,i}$ can be represented as
\begin{equation}\label{eq:pder_wrt_nu}
    \frac{\partial \v{W}}{\partial \nu_{l,i}} = \v{P}_{\v{E}_l}\v{A}\v{e}_i\v{e}_i^T
\end{equation}
where 
\begin{meq}
    \v{P}_{\v{E}_l} & = \v{E}_l\v{E}_l^T,~\v{E}_l = \begin{bmatrix}
                    \v{0}_{a_l\times N_l}^T&
                    \v{I}_{N_l\times N_l}&
                    \v{0}_{b_l\times N_l}^T\\
                \end{bmatrix}^T\\    
    a_l & = \sum_{i=1}^{l-1}N_i,~ b_l = \sum_{i=l+1}^{L}N_i\\
    \v{A} & = \sqb{\v{A}_{\set{S}_1}^T(\v{\theta}),\ldots,\v{A}_{\set{S}_L}^T(\v{\theta})}^T\\
\end{meq}

By substituting (\ref{eq:pder_wrt_nu}) in (\ref{eq:der_nu_first}), we can write
\begin{meq}\label{eq:der_wrt_nu}
    \frac{\partial \v{R}_{\set{S}}(\v{\phi})}{\partial \nu_{l,i}} & = \v{P}_{\v{E}_l}\v{A}\v{e}_i\v{e}_i^T\v{R}_s\v{W}^H+\v{W}\v{R}_s\v{e}_i\v{e}_i^T\v{A}^H\v{P}_{\v{E}_l}\\
\end{meq}

\item Derivatives with respect to the imaginary part of the calibration parameters $\v{\eta}_l$

Due to the structure and relation between $\v{\nu}_l$ and $\v{\eta}_l$, the derivatives $\partial \v{W}/\partial \eta_{l,i}$ follow immediately from (\ref{eq:pder_wrt_nu}) as
\begin{meq}\label{eq:wder_wrt_eta}
\frac{\partial \v{W}}{\partial \eta_{l,i}} & = j\frac{\partial \v{W}}{\partial \nu_{l,i}}\\
            & = j\v{P}_{\v{E}_l}\v{A}\v{e}_i\v{e}_i^T\\
\end{meq}
The derivative of the covariance matrix with respect to $\eta_{l,i}$ is given by
\begin{equation}\label{eq:gen_deriv_eta}
    \frac{\partial \v{R}_{\set{S}}(\v{\phi})}{\partial \eta_{l,i}} = \frac{\partial \v{W}}{\partial \eta_{l,i}}\v{R}_s\v{W}^H+\v{W}\v{R}_s\frac{\partial \v{W}^H}{\partial \eta_{l,i}}\\
\end{equation}
By replacing (\ref{eq:wder_wrt_eta}) into (\ref{eq:gen_deriv_eta}) results in
\begin{meq}\label{eq:der_wrt_eta}
    \frac{\partial \v{R}_{\set{S}}(\v{\phi})}{\partial \eta_{l,i}} & = j\left(\v{P}_{\v{E}_l}\v{A}\v{e}_i\v{e}_i^T\v{R}_s\v{W}^H-\right.\\
    &\left. \v{W}\v{R}_s\v{e}_i\v{e}_i^T\v{A}^H\v{P}_{\v{E}_l}\right)\\
\end{meq}

\end{itemize}

\noindent
II) Calculation of the FIM matrices
\begin{enumerate}

\item FIM matrix with respect to the DOAs $\v{\theta}$

By substituting (\ref{eq:deriv_theta}) into (\ref{eq:gen_fim_formula}), we have
\begin{meq}
    &   \sqb{\text{FIM}_{\v{\theta\theta}}}_{ij} = \\
    &   T\opn{tr}\{\v{R}_{\set{S}}^{-1}\v{D}\v{e}_i\v{e}_i^T\v{R}_s\v{W}^H\v{R}_{\set{S}}^{-1}\v{D}\v{e}_j\v{e}_j^T\v{R}_s\v{W}^H\\
    & + \v{R}_{\set{S}}^{-1}\v{D}\v{e}_i\v{e}_i^T\v{R}_s\v{W}^H\v{R}_{\set{S}}^{-1}\v{W}\v{R}_s\v{e}_j\v{e}_j^T\v{D}^H\\
    & + \v{R}_{\set{S}}^{-1}\v{WR}_s\v{e}_i\v{e}_i^T\v{D}^H\v{R}_{\set{S}}^{-1}\v{D}\v{e}_j\v{e}_j^T\v{R}_s\v{W}^H\\
    & + \v{R}_{\set{S}}^{-1}\v{WR}_s\v{e}_i\v{e}_i^T\v{D}^H\v{R}_{\set{S}}^{-1}\v{W}\v{R}_s\v{e}_j\v{e}_j^T\v{D}^H\}
 \end{meq} 
By applying the cyclic property of the trace, we obtain:
\begin{meq}
    & \sqb{\text{FIM}_{\v{\theta\theta}}}_{ij} = \\
    & T\opn{tr}\{(\v{e}_j^T\v{R}_s\v{W}^H\v{R}_{\set{S}}^{-1}\v{D}\v{e}_i)(\v{e}_i^T\v{R}_s\v{W}^H\v{R}_{\set{S}}^{-1}\v{D}\v{e}_j)\\
    & + (\v{e}_j^T\v{R}_s\v{W}^H\v{R}_{\set{S}}^{-1}\v{WR}_s\v{e}_i)(\v{e}_i^T\v{D}^H\v{R}_{\set{S}}^{-1}\v{D}\v{e}_j)\\
    & +(\v{e}_j^T\v{D}^H\v{R}_{\set{S}}^{-1}\v{D}\v{e}_i)(\v{e}_i^T\v{R}_s\v{W}^H\v{R}_{\set{S}}^{-1}\v{WR}_s\v{e}_j)\\
    & +(\v{e}_j^T\v{D}^H\v{R}_{\set{S}}^{-1}\v{WR}_s\v{e}_i)(\v{e}_i^T\v{D}^H\v{R}_{\set{S}}^{-1}\v{WR}_s\v{e}_j)\}
 \end{meq}   
 which can be rewritten without the trace as:
 \begin{meq}   
    & \sqb{\text{FIM}_{\v{\theta\theta}}}_{ij} = \\
    & 2T\mathfrak{Re}\{(\v{e}_i^T\v{R}_s\v{W}^H\v{R}_{\set{S}}^{-1}\v{D}\v{e}_j)(\v{e}_j^T\v{R}_s\v{W}^H\v{R}_{\set{S}}^{-1}\v{D}\v{e}_i)\\
    & +(\v{e}_i^T\v{R}_s\v{W}^H\v{R}_{\set{S}}^{-1}\v{WR}_s\v{e}_j)(\v{e}_j^T\v{D}^H\v{R}_{\set{S}}^{-1}\v{D}\v{e}_i)\}
\end{meq}
or in matrix form
\begin{meq}
    & \text{FIM}_{\v{\theta\theta}} = \\
    & 2T\mathfrak{Re}\{(\v{R}_s\v{W}^H\v{R}_{\set{S}}^{-1}\v{D})\odot (\v{R}_s\v{W}^H\v{R}_{\set{S}}^{-1}\v{D})^T\\
    &+(\v{R}_s\v{W}^H\v{R}_{\set{S}}^{-1}\v{WR}_s)\odot(\v{D}^H\v{R}_{\set{S}}^{-1}\v{D})^T\}\\
\end{meq}
where $\odot$ denotes the Hadamard product. Due to space constraints, only the final results of the subsequently computed FIMs are presented.

\item FIM matrix with respect to the sources powers $\v{p}$

By replacing (\ref{eq:deriv_sources_powers}) into (\ref{eq:gen_fim_formula}), we can write
\begin{equation}
    \text{FIM}_{\v{p}\v{p}} = T(\v{W}^H\v{R}_{\set{S}}^{-1}\v{W})\odot(\v{W}^H\v{R}_{\set{S}}^{-1}\v{W})^T 
\end{equation}


\item FIM matrix with respect to the noise power $\sigma_n^2$

By substituting (\ref{eq:deriv_noise_power}) into (\ref{eq:gen_fim_formula}), we arrive at
\begin{equation}   
    \text{FIM}_{\sigma_n^2\sigma_n^2} = T\opn{tr}\sqb{\v{R}_{\set{S}}^{-2}} 
\end{equation}

\item FIM matrix with respect to the real part of the calibration parameters $\v{\nu}_l$

By substituting (\ref{eq:der_wrt_nu}) into (\ref{eq:gen_fim_formula}), we can write

\begin{meq}\label{eq:fim_mat_nu}
    & \text{FIM}_{\v{\nu}_l\v{\nu}_k} = \\
    & 2T\mathfrak{Re}\{(\v{R}_s\v{W}^H\v{R}_{\set{S}}^{-1}\v{P}_{\v{E}_k}\v{A})\odot(\v{R}_s\v{W}^H\v{R}_{\set{S}}^{-1}\v{P}_{\v{E}_l}\v{A})^T\\
    & +(\v{R}_s\v{W}^H\v{R}_{\set{S}}^{-1}\v{W}\v{R}_s)\odot(\v{A}^H\v{P}_{\v{E}_k}\v{R}_{\set{S}}^{-1}\v{P}_{\v{E}_l}\v{A})^T\}
\end{meq}

\item FIM matrix with respect to the imaginary part of the calibration parameters $\v{\eta}_l$

By substituting (\ref{eq:der_wrt_eta}) into (\ref{eq:gen_fim_formula}), we can write

\begin{meq}\label{eq:fim_mat_eta}
    & \text{FIM}_{\v{\eta}_l\v{\eta}_k} = \\
    & -2T\mathfrak{Re}\{(\v{R}_s\v{W}^H\v{R}_{\set{S}}^{-1}\v{P}_{\v{E}_k}\v{A})\odot(\v{R}_s\v{W}^H\v{R}_{\set{S}}^{-1}\v{P}_{\v{E}_l}\v{A})^T\\
    & -(\v{R}_s\v{W}^H\v{R}_{\set{S}}^{-1}\v{W}\v{R}_s)\odot(\v{A}^H\v{P}_{\v{E}_k}\v{R}_{\set{S}}^{-1}\v{P}_{\v{E}_l}\v{A})^T\}\\
\end{meq}
\item FIM matrix with respect to  $\v{\theta}$ and $\v{p}$
\begin{equation}\label{eq:fim_theta_p_final}
\text{FIM}_{\v{\theta}\v{p}} = 2T\mathfrak{Re}\{(\v{R}_s\v{W}^H\v{R}_{\set{S}}^{-1}\v{W})\odot(\v{W}^H\v{R}_{\set{S}}^{-1}\v{D})^T\}
\end{equation}

\item FIM matrix with respect to  $\v{\theta}$ and $\sigma_n^2$
\begin{equation}\label{eq:fim_theta_noise_second}
\text{FIM}_{\v{\theta}\sigma_n^2} = 2T\opn{diag}^M(\mathfrak{Re}\left\{\v{R}_s\v{W}^H\v{R}_{\set{S}}^{-2}\v{D}\right\})
\end{equation}
where $\opn{diag}^M(\v{X})$ is a column vector whose elements consists of the diagonal elements of square-matrix $\v{X}$.

\item FIM matrix with respect to  $\v{\theta}$ and $\v{\nu}_l$
\begin{meq}\label{eq:fim_theta_nu_second}
& \text{FIM}_{\v{\theta}\v{\nu}_l} = \\
& = T\{(\v{R}_s\v{W}^H\v{R}_{\set{S}}^{-1}\v{P}_{\v{E}_l}\v{A})\odot (\v{R}_s\v{W}^H\v{R}_{\set{S}}^{-1}\v{D})^T\\
& +(\v{R}_s\v{W}^H\v{R}_{\set{S}}^{-1}\v{W}\v{R}_s)\odot(\v{A}^H\v{P}_{\v{E}_l}\v{R}_{\set{S}}^{-1}\v{D})^T\\
& +(\v{D}^H\v{R}_{\set{S}}^{-1}\v{P}_{\v{E}_l}\v{A})\odot(\v{R}_s\v{W}^H\v{R}_{\set{S}}^{-1}\v{W}\v{R}_s)^T\\
& +(\v{D}^H\v{R}_{\set{S}}^{-1}\v{W}\v{R}_s)\odot(\v{A}^H\v{P}_{\v{E}_l}\v{R}_{\set{S}}^{-1}\v{W}\v{R}_s)^T\}\\
\end{meq}

\item FIM matrix with respect to  $\v{\theta}$ and $\v{\eta}_l$

\begin{meq}\label{eq:fim_theta_eta_second}
& \text{FIM}_{\v{\theta}\v{\eta}_l} = \\
& = jT\{(\v{R}_s\v{W}^H\v{R}_{\set{S}}^{-1}\v{P}_{\v{E}_l}\v{A})\odot(\v{R}_s\v{W}^H\v{R}_{\set{S}}^{-1}\v{D})^T\\
& -(\v{R}_s\v{W}^H\v{R}_{\set{S}}^{-1}\v{W}\v{R}_s)\odot(\v{A}^H\v{P}_{\v{E}_l}\v{R}_{\set{S}}^{-1}\v{D})^T\\
& +(\v{D}^H\v{R}_{\set{S}}^{-1}\v{P}_{\v{E}_l}\v{A})\odot(\v{R}_s\v{W}^H\v{R}_{\set{S}}^{-1}\v{W}\v{R}_s)^T\\
& -(\v{D}^H\v{R}_{\set{S}}^{-1}\v{W}\v{R}_s)\odot(\v{A}^H\v{P}_{\v{E}_l}\v{R}_{\set{S}}^{-1}\v{W}\v{R}_s)^T\}\\
\end{meq}

\item FIM matrix with respect to  $\v{p}$ and $\v{\sigma}_n^2$

\begin{equation}
    \text{FIM}_{\v{p}\sigma_n^2} = T\opn{diag}^M(\v{W}^H\v{R}_{\set{S}}^{-2}\v{W})
\end{equation}

\item FIM matrix with respect to  $\v{p}$ and $\v{\nu}_l$

\begin{meq}\label{eq:fim_p_nu_second}
& \text{FIM}_{\v{p}\v{\nu}_l} = \\
& = 2T\mathfrak{Re}\left\{\prt{\v{W}^H\v{R}_{\set{S}}^{-1}\v{P}_{\v{E}_l}\v{A}}\odot \prt{\v{R}_s\v{W}^H\v{R}_{\set{S}}^{-1}\v{W}}^T\right\}\\
\end{meq}

\item FIM matrix with respect to  $\v{p}$ and $\v{\eta}_l$

\begin{meq}\label{eq:fim_p_eta_second}
& \text{FIM}_{\v{p}\v{\eta}_l} = \\
& -2T\mathfrak{Im}\{(\v{W}^H\v{R}_{\set{S}}^{-1}\v{P}_{\v{E}_l}\v{A})\odot(\v{R}_s\v{W}^H\v{R}_{\set{S}}^{-1}\v{W})^T\}\\
\end{meq}

\item FIM matrix with respect to  $\sigma_n^2$ and $\v{\nu}_l$
\begin{meq}\label{eq:fim_n_nu_second}
& \text{FIM}_{\sigma_n^2\v{\nu}_l} = 2T\opn{diag^{M}}(\mathfrak{Re}\{\v{R}_s\v{W}^H\v{R}_{\set{S}}^{-2}\v{P}_{\v{E}_l}\v{A}\})^T\\
\end{meq}

\item FIM matrix with respect to $\sigma_n^2$ and $\v{\eta}_l$
\begin{meq}\label{eq:fim_n_eta_second}
& \text{FIM}_{\sigma_n^2\v{\eta}_l} = -2T\opn{diag^{M}}(\mathfrak{Im}\{\v{R}_s\v{W}^H\v{R}_{\set{S}}^{-2}\v{P}_{\v{E}_l}\v{A}\})^T\\
\end{meq}

\item FIM matrix with respect to $\v{\nu}_l$ and $\v{\eta}_k$

\begin{meq}
    & \text{FIM}_{\v{\nu}_l\v{\eta}_k} = \\
    & -2T\mathfrak{Im}\{(\v{R}_s\v{W}^H\v{R}_{\set{S}}^{-1}\v{P}_{\v{E}_k}\v{A})\odot(\v{R}_s\v{W}^H\v{R}_{\set{S}}^{-1}\v{P}_{\v{E}_l}\v{A})^T\\
    & +(\v{A}^H\v{P}_{\v{E}_l}\v{R}_{\set{S}}^{-1}\v{P}_{\v{E}_k}\v{A})\odot (\v{R}_s\v{W}^H\v{R}_{\set{S}}^{-1}\v{W}\v{R}_s)^T\}\\
 \end{meq}

 \end{enumerate}

Since the general FIM is symmetric, the above expressions are exhaustive in describing the amount of information from all the model parameters embedded in the data. By taking the FIM inverse using (\ref{eq:crm_phi1_phi1}), we have the CRLB for the DOAs. However, beyond yielding the CRLB for DOA estimates, these expressions can be used to calculate the CRLB for any parameter in the data model (\ref{eq:data_model_param}). 

They can even be used for the calculus of the CRLB of the fully calibrated array if we set $L = 1$ and remove the rows and columns corresponding to the calibration parameters. Besides, they do not have any limitations with the quantity of sensors in relation to the number of sources and consider the uncorrelated prior hypothesis for the sources, which is a core assumption for algorithms like GCA-MUSIC and GCA-rMUSIC.

\section{Simulation Results and Discussion}\label{sec:results}

In this section, we employ simulations to assess the algorithms' capabilities and demonstrate the applicability of the CRLB developed in this work. The performance curves are shown in terms of root mean squared-error (RMSE) in two distinct ways: comparison of the algorithms for a given geometry and comparison of the geometries for a given algorithm. All the results were obtained from an average of 50,000 simulation trials. 
In what follows, we outline the parameter setting, emphasizing that the number of signal sources surpasses the number of sensor elements within each subarray.
\begin{enumerate}
  \item Number of snapshots: $T=2000$ (for the RMSE against SNR plots)
  \item Number of subarrays: $L=2$ 
  \item $\text{SNR} = 0$ dB (for the RMSE against T plots)
  \item Type-II array with $N_l=7$ sensors (total of $2N_l=14$ sensors) and the following geometries for the reference subarray: a) II-MRA: $\set{S}_l=\{0,1,4,10,12,15,17\}$; 
         b) II-SNAQ2: $\set{S}_l=\{0,2,3,6,9,13,14\}$; c) II-NAQ2: $\set{S}_l=\{0,1,2,3,4,9,14\}$
  \item Normalized intersubarray displacement: $\mu=8$, which implies that $\v{h}(\theta_d)=[1,\exp(j\pi(\mu+\kappa)\theta_d)]^T$ corresponds to the vector of calibration parameters for $\theta_d = \sqb{\v{\theta}}_d$
  \item $D=11$ sources with directions given by: $\v{\theta}=\sqb{\pm 0.75,\pm 0.6, \pm 0.45, \pm 0.3, \pm 0.15, 0}^T$
\end{enumerate}

\subsection{Analysis of degrees of freedom of full array and subarrays}

\begin{figure}
  \centering
  \includegraphics[width=.47\textwidth]{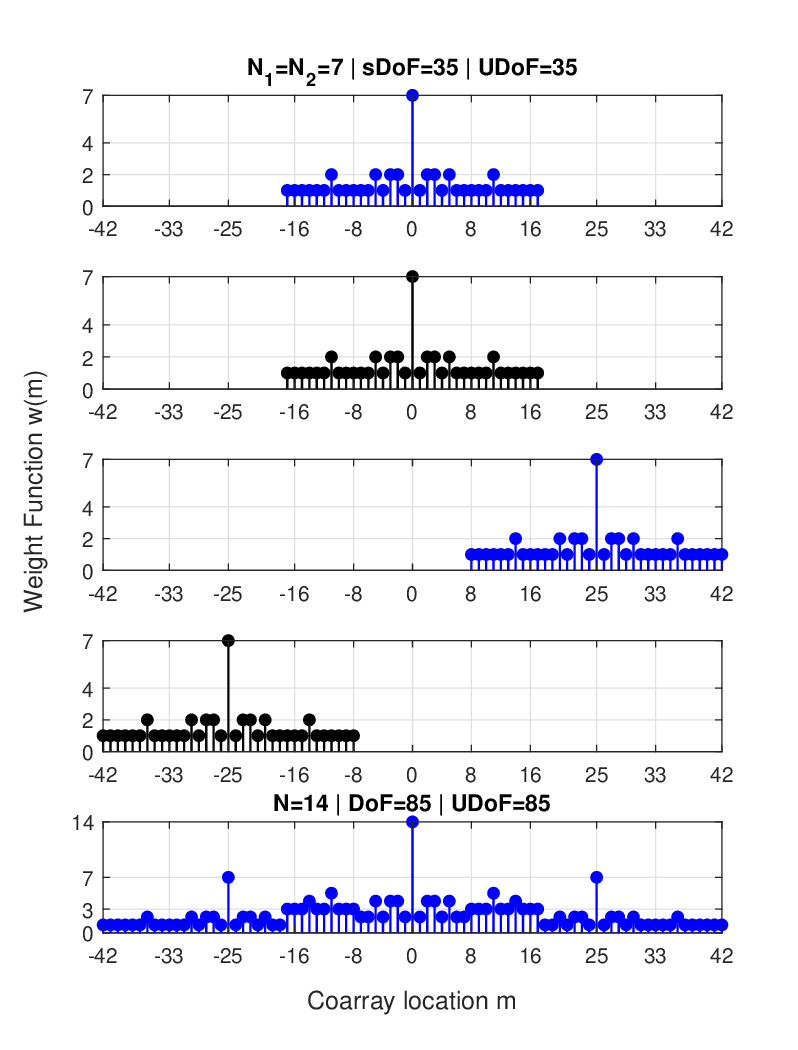} \vspace{-1em}
  \caption{Weight function of the full array as a result of the weight function of subarrays for II-MRA.}\label{fig:weight_func}
\end{figure}

In this section, we aim to characterize the weight function of the full array based on the weight function of the subarrays for Type II-MRA. To achieve this, we use the results from Theorem~\ref{thm:dofsdof}, derived in Section~\ref{sec:anadof}. We observe that the intersubarray displacement is denoted by $\mu=8$ and the aperture of each subarray is given by $\kappa=17-0=17$. Additionally, the computation of the weight functions of the subarrays, as shown in the first two subplots of Fig.~\ref{fig:weight_func}, indicates that the subarrays DoF and UDoF are equal to $\textrm{sDoF}=\textrm{UDoF}=35$. Consequently, Theorem~\ref{thm:dofsdof} dictates that the total DoF of the full array is given by 
\begin{meq}
    \textrm{DoF} & = L(\text{sDoF}-1)+2(L-1)\mu+1\\
                 & = 2\cdot(35-1)+2(2-1)8+1=85
\end{meq}
which agrees with the number of non-zero elements in the weight function of the entire array, as depicted in the fifth subplot of Fig.~\ref{fig:weight_func}. It is also important to note also that this final weight function represents the sum of the first four weight functions. The weight functions shown in the third and fourth subplots are simply the weight functions of the subarrays translated by $\Delta=\pm(\mu+\kappa)=\pm 25$, as expected.

\subsection{Computational complexity}

The primary objective of GCA-(r)MUSIC is to perform DOA estimation when the number of sources is greater than that of sensors under two distinct scenarios: a) in fully calibrated arrays, where we advocate for non-coherent data processing to reduce hardware complexity and computational costs; and (b) in scenarios where calibration parameters remain unidentified, recommending the use of these algorithms to capitalize on their capability to estimate the DOAs with high performance, outperforming existing algorithms tailored for partially-calibrated scenarios. Regarding computational complexity, our approach is benchmarked in the first scenario against the SS-MUSIC (or coarray MUSIC) algorithm \cite{Pal2010}.

For the subsequent scenario, we draw a comparison with Spectral RARE and G-MUSIC, tailored for partially-calibrated arrays \cite{See2004,Rieken2004}. Notably, while other algorithms in the literature address partially-calibrated arrays, they were excluded from our comparison due to their applicability solely to ULAs \cite{Suleiman2014} or because their underlying signal model presupposes that the cross-covariance between subarrays is zero \cite{suleiman2018}, a condition not applicable in our context.

In the subsequent simulation, we utilized a simulation server with 128 GB of RAM and an Intel Core i9-10980XE CPU at 3.0 GHz. The coding for all methods was optimally performed to minimize the number of required flops as much as possible. The normalized spacing between subarrays was assumed known for the SS-MUSIC algorithm (calibrated scenarios only) and unknown for the other four algorithms (GCA-(r)MUSIC, G-MUSIC and Spectral RARE). The spacing between subarrays constitutes our calibration information. The results concerning runtime, averaged over 50,000 simulation trials, are presented in Table~\ref{tab:comp_complexity}.

GCA-MUSIC and GCA-rMUSIC operate much faster than the other three methods. Moreover, G-MUSIC fails to identify the sources due to identifiability issues when the number of sources exceeds the number of sensors within a subarray. We also highlight that SS-MUSIC is approximately 12 times slower than the proposed approach and only can be used if all of the calibration parameters are known (in our case, the distance between subarrays). Despite those limitations, with a much higher computational and required hardware complexity, it provides the lowest RMSE because it does the processing using the calibration parameters, as expected. Among the algorithms designed for partially-calibrated scenarios, GCA-rMUSIC and GCA-MUSIC run approximately 4.3 and 3.3 times faster than Spectral MUSIC, respectively.       

\begin{table}
  \caption{Runtime of GCA-({r})MUSIC against established approaches.} \vspace{-1em} \label{tab:comp_complexity} 
  \centering
  \begin{tabular}{c|c|c}
    \hline\hline
    Algorithm & Runtime (Norm. Runtime) & DOA RMSE\\
    \hline
    GCA-rMUSIC & $1.61$~ms (1.00) & 1.0943e-3\\
    \hline
    GCA-MUSIC & $2.10$~ms(1.306) & 1.0954e-3\\
    \hline
    SS-MUSIC & $12.15$~ms (7.54) & 0.15e-3\\
    \hline
    Spectral RARE & $6.96$~ms (4.32) & 8.45e-3\\
    \hline 
    G-MUSIC & $2.45$~ms (1.52) & 0.99629\\
    \hline\hline
  \end{tabular}
\end{table}

\subsection{CRLB comparison}

Regarding the CRLB, we consider five bounds:
\begin{enumerate}
    \item CRLB-FC: fully calibrated array without any special structure for the covariance matrix of the sources \cite{Stoica1989};
    \item CRLB-FC-UP: fully calibrated array assuming the sources are uncorrelated \emph{a priori} - (\emph{Uncorrelated Prior}), originally obtained in \cite{Jansson1999} and extended to deal with sparse arrays for the underdetermined case in \cite{Liu2017}. It assumes the covariance matrix of the sources is diagonal;
    \item CRLB-PC: partially-calibrated array with covariance matrix of the sources without any special structure \cite{See2004};
    \item \textbf{CRLB-PC-UP-PROP}: partially-calibrated array assuming the sources are uncorrelated \emph{a priori} (covariance matrix of the sources with diagonal structure), which is our \emph{proposed bound}; and
    \item CRLB-PC-UXCov: partially-calibrated array assuming the cross-covariances between subarrays are zero-valued, i.e., the sources impinging on different subarrays are uncorrelated, proposed in \cite{suleiman2018}. 
\end{enumerate}

The proposed CRLB-PC-UP-PROP bound is a unique contribution to the literature, providing a lower bound for the estimation performance of DOAs. This CRLB uniquely incorporates two critical assumptions: the uncorrelated nature of the sources, which is a fundamental consideration in coarray MUSIC algorithms, and the partial calibration of arrays. Unlike existing bounds such as CRLB-PC, CRLB-FC, and CRLB-PC-UXCov, the proposed bound overcomes numerical challenges encountered when the number of sources exceeds the number of sensors in the full array. This distinction not only underscores the practicality of our approach but also its applicability in scenarios often encountered in array signal processing.

\begin{figure}
  \centering
  \includegraphics[width=.47\textwidth]{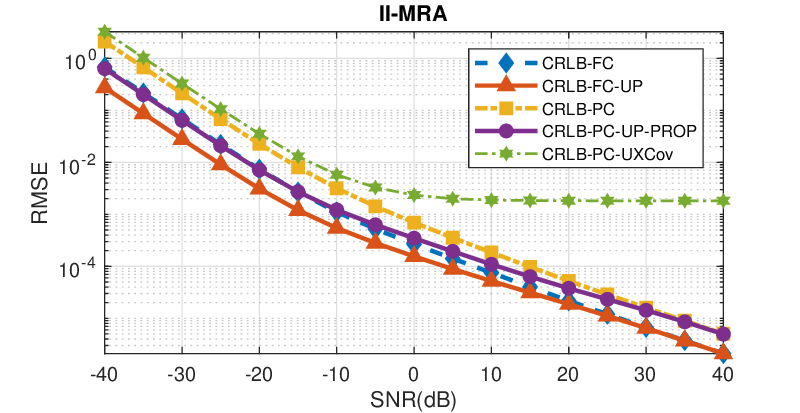} \vspace{-1em}
  \caption{CRLB curves against SNR for type-II MRA.}\label{fig:crlb_snr_comp}
\end{figure}

The CRLBs above are depicted in Fig.~\ref{fig:crlb_snr_comp}. Among these bounds, it is evident that those assuming uncorrelated sources (appended by ``-UP''), although initially lower, converge to the same value as its counterpart as the SNR increases. This convergence is observed in the pairs CRLB-FC/CRLB-FC-UP and CRLB-PC/CRLB-PC-UP-PROP. This phenomenon indicates that the uncertainty in the off-diagonal parameters of the sources' covariance matrix become negligible for the DOA estimation performance in substantially high SNR scenarios. Note that CRLB-PC-UP-PROP differs significantly from CRLB-PC for SNR values below $10$~dB.

We also emphasize that CRLB-FC-UP is the bound with the lowest values, as it corresponds to a scenario with a fully calibrated array and uncorrelated sources, resulting in reduced uncertainty for DOA estimation algorithms. Lower uncertainty leads to better performance in estimation procedures. Moreover, we note that CRLB-PC-UXCov is the highest bound and, contrary to the other CRLB curves, it reaches a plateau after $10$ dB. This behaviour, as expected and explained in \cite{suleiman2018}, is due to the subarrays' covariance matrices not being full rank (i.e., more sources than sensors in each subarray). It is also important to mention that by employing the FIM expressions in Section~\ref{sec:deriv_crlb}, we can estimate the CRLB for all parameters of the data model, not only the DOAs. This is because these matrices are explicitly provided, which extends the applicability of the work presented here to many parameter estimation algorithms.
  
\begin{figure}
  \centering
  \includegraphics[width=.47\textwidth]{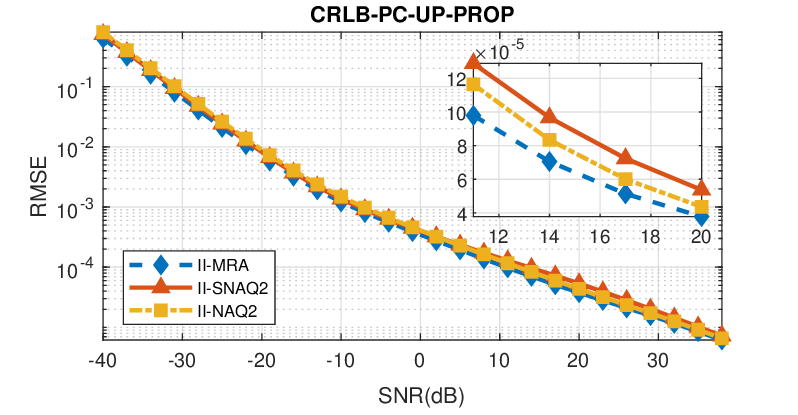} \vspace{-1em}
  \caption{Proposed CRLB-PC-UP-PROP against SNR for II-MRA, II-SNAQ2 and II-NAQ2.}\label{fig:crlb_snr_comp_zoom}
\end{figure}

In Fig.~\ref{fig:crlb_snr_comp_zoom}, we present a comparison of the three type-II array geometries in terms of the CRLB using CRLB-PC-UP-PROP. The curves, plotted against SNR, suggest that DOA estimation algorithms are likely to perform better with the type II-MRA geometry. This geometry consistently exhibits the lowest CRLB values. This theoretical aspect is corroborated by practical outcomes, as will be evidenced in Section~\ref{sec:rmse_comp}. To allow a clearer comparison, we focused on the SNR range from $10$~dB to $20$~dB, displaying the curves in the right upper squared box in the figure.

\subsection{RMSE performance comparison}\label{sec:rmse_comp}

Before we enter into the RMSE discussion, it is worth exploring the expected RMSE for this scenario. Consider that a given estimation procedure obtains estimates of a random angle $\hat{\theta}_d$ from $(-1,1]$. Additionally, suppose that this random variable follows a uniform distribution  $\hat{\theta}_d\sim \mathcal{U}(-1,1)$. Then, the RMSE for this naive algorithm would be given by
$\text{RMSE}^\text{naive} = \sqrt{\frac{1}{3}+\frac{1}{D}\sum_{d=1}^{D}\theta_d^2}$. Indeed, for this scenario we have $\text{RMSE}^\text{naive}=0.7472$. Then, for an RMSE smaller than this, the estimator is better than a random choice for the DOAs. This is useful as a benchmark for the challenging scenarios in terms of data availability and SNR. Although it is not formally an upper bound on the estimator's RMSE, it is a metric we should pay attention to.

\begin{figure}
  \centering
  \includegraphics[width=.47\textwidth]{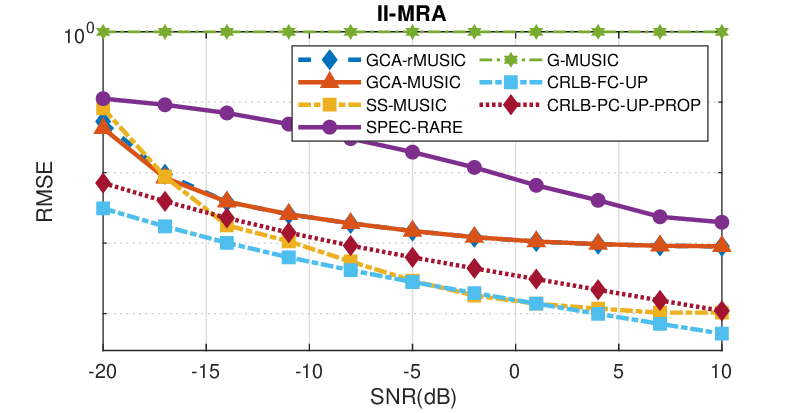} \vspace{-1em}
  \caption{Algorithm comparison. RMSE curves against SNR. GCA-MUSIC and GCA-rMUSIC for type-II arrays.}\label{fig:rmse_alg_snr}
\end{figure}

\begin{figure}
  \centering
  \includegraphics[width=.47\textwidth]{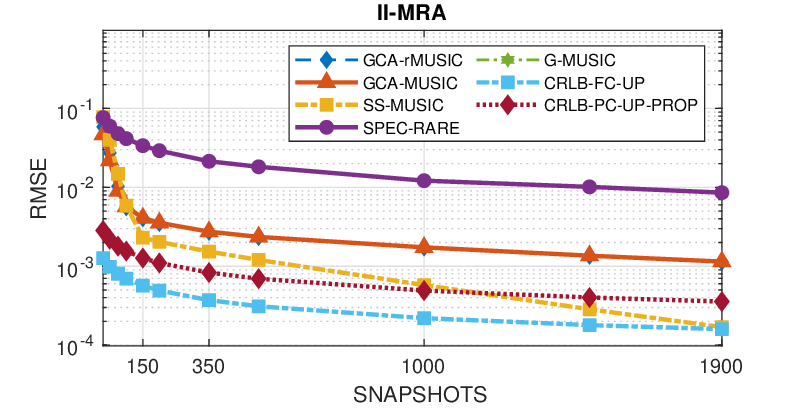} \vspace{-1em}
  \caption{Algorithm comparison. RMSE curves against SNAPSHOTS. GCA-MUSIC and GCA-rMUSIC for type-II arrays.}\label{fig:rmse_alg_snp}
\end{figure}

Next, we present the RMSE curves in Fig.~\ref{fig:rmse_alg_snr}, where the proposed algorithms are able to identify the sources in scenarios with more sources than sensors in each subarray and that GCA-rMUSIC slightly outperforms GCA-MUSIC for the lower SNR values. Although SS-MUSIC presents the best performance, it only applies to calibrated schemes, has a higher computational burden and we show its RMSE only for comparison purposes. The CRLB curve of our data model, as it was derived in Section \ref{sec:deriv_crlb}, CRLB-PC-UP-PROP shows that our algorithms get closer to optimum in the interval $-17$ dB up to $-5$ dB, saturating the performance as the SNR increases, which agrees with the analysis developed in the fully calibrated case for coarray MUSIC \cite{wang2017}. The RMSE value of $10^{-2}$ shows that Spectral RARE only achieves the performance of the proposed algorithms after the SNR increases by 15 dB. Additionally, we also observe that G-MUSIC fails to estimate the DOA of the sources because the number of sources exceeds the number of sensors in each subarray. Note that the optimization-based algorithms from \cite{Tirer2020} and \cite{suleiman2018} were excluded from our comparisons due to their significantly higher computational complexity—convex optimization iterative schemes without closed-form solutions for the former, 
and assumptions of zero-valued cross-covariance between subarrays for the latter. Moreover, these algorithms are based on data models that are different from the one used in this paper. Similar conclusions can be inferred from the RMSE curves versus the snapshots in Fig.~\ref{fig:rmse_alg_snp}.

\begin{figure}
  \centering
  \includegraphics[width=.47\textwidth]{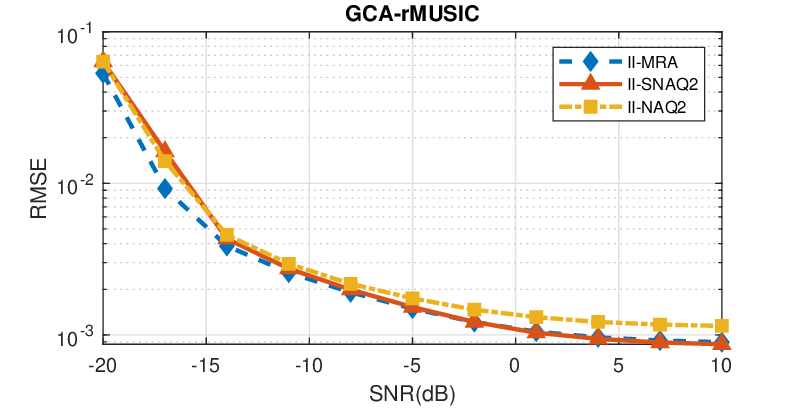} \vspace{-1em}
  \caption{Geometry comparison for GCA-root MUSIC. RMSE curves against SNR. Type-II MRA, NAQ2 and SNAQ2.}\label{fig:rmse_geom_rmusic}
\end{figure}

\begin{figure}
  \centering
  \includegraphics[width=.47\textwidth]{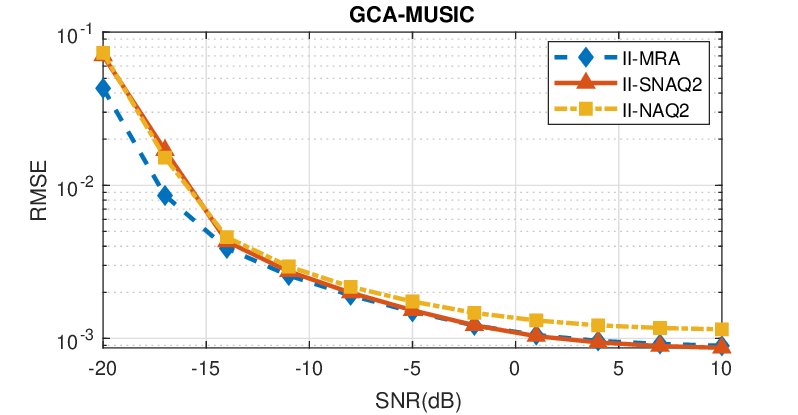} \vspace{-1em}
  \caption{Geometry comparison for GCA-MUSIC. RMSE curves against SNR. Type-II MRA, NAQ2 and SNAQ2.}\label{fig:rmse_geom_music}
\end{figure}

In Fig.~\ref{fig:rmse_geom_rmusic} and Fig.~\ref{fig:rmse_geom_music}, we can verify that the performance varies largely with the geometry for both algorithms and that II-MRA presents the best estimation performance due to its larger number of DoF (virtual aperture), according to what was foreseen from the analysis of Fig.~\ref{fig:crlb_snr_comp_zoom} in the previous section. The performance of II-MRA is achieved by II-SNAQ2 only after the thresholding of $-5$ dB for the SNR range.

\section{Conclusions}\label{sec:conclusions}

In this study, we have developed and analyzed novel design techniques for partially-calibrated sparse linear subarrays and introduced the algorithms GCA-MUSIC and GCA-rMUSIC for DOA estimation. These estimators effectively address the limitations of existing DOA estimation techniques, particularly in underdetermined scenarios where the number of sources exceeds the number of sensors. The proposed algorithms leverage the concept of sparse subarray design within the coarray domain to enhance estimation accuracy while maintaining computational efficiency. This approach not only expands the applicability of coarray-based signal processing to partially-calibrated arrays but also significantly increases the number of sources that can be estimated accurately. Through rigorous analysis, including the computation of the CRLB and extensive simulations, our methods have been shown to outperform existing approaches in terms of estimation precision and resource efficiency. This has the potential to impact various applications in sensor array processing by offering a more flexible and practical solution to the DOA estimation problem.

\bibliographystyle{IEEEtran}
\bibliography{mybib}

\end{document}